\documentclass[12pt,english,twoside]{article}

\usepackage{AK}
\usepackage{amsfonts}
\usepackage{amssymb}
\usepackage{amsmath}
\usepackage{newlfont}
\usepackage{dsfont}
\usepackage[longnamesfirst,sort]{natbib}
\usepackage[svgnames]{xcolor}
\usepackage{graphicx}
\usepackage[colorlinks=true,bookmarks=false,implicit=false,urlcolor=darkblue]{hyperref}
\usepackage{qrcode}

\bibpunct{(}{)}{,}{a}{,}{,}
\newcommand{\SR}{{\rm I\kern-.21em R}}

\newcommand{\eenn}{\end{eqnarray}}

\def\player(#1,#2){\langle#1,#2\rangle}

\newcommand\blfootnote[1]{%
  \begingroup
  \renewcommand\thefootnote{}\footnote{#1}%
  \addtocounter{footnote}{-1}%
  \endgroup
}
\input tcilatex
\begin{document}
\definecolor{darkblue}{rgb}{0,0,0.35} 
\title{\vspace{-1em}
 \fontsize{32}{30}\selectfont{Approximately Optimal Auctions With a Strong Bidder$^{*}$}}
\author{\hspace{0em}\href{http://www.anderlini.net}{Luca Anderlini}\\
\hspace{0em}\begin{tabular}{c}\textit{\href{http://www.georgetown.edu}{Georgetown University} and}\\
\textit{\href{http://www.unina.it/en_GB/home}{University of Naples Federico II}}
\end{tabular}
\and \href{https://mitsloan.mit.edu/programs/phd/gaon-kim}{GaOn Kim}\\
\textit{\href{https://mitsloan.mit.edu}{MIT Sloan School of Management}}
\\
 \vspace{0.45em}
}

\date{\vspace{1.5em}\data}

\maketitle
\keywords{Strong Insider, Tournament Auction, Approximate Optimality\hspace{-0.3em}
\vspace{0.2em}
 } 

\jel{\href{http://www.aeaweb.org/econlit/jelCodes.php?view=econlit}{C70},
\href{http://www.aeaweb.org/econlit/jelCodes.php?view=econlit}{C72},
\href{http://www.aeaweb.org/econlit/jelCodes.php?view=econlit}{C79.}$\,$\hspace{-0.8em}
\vspace{0.2em}
 } 

\simplecorrespondent{\href{http://anderlini.net}{Luca Anderlini} $\;$ --- $\;$
\href{mailto:luca@anderlini.net}{{\tt luca@anderlini.net}}. Department of Economics, Georgetown University,
$37^{th}$ and O Streets, Washington DC 20057, USA.}

\runninghead{\textsc{Approximately Optimal Auctions}}

\runningauthor{\textsc{Anderlini and Kim}} 

\definecolor{darkblue}{rgb}{0,0,0.35} %
\vspace{1em}

\begin{abstract}\hspace{-0.5em}We consider auctions with $N+1$ bidders. Of these, $N$ are
symmetric and $N+1$ is  ``sufficiently strong'' relative to the others. 
The auction is a ``tournament'' in which the first $N$ players bid to win the right to compete
with $N+1$. The bids of the first $N$ players are binding and the highest bidder proceeds to a second-price competition with $N+1$.

When $N+1$'s values converge in distribution to an atom above the upper end of
the distribution of the $N$ bidders and the rest of the 
distribution is drained away from low values sufficiently slowly, the auction's
expected revenue is arbitrarily close to the one obtained in a \citet{Myerson:81} optimal auction.

The tournament design is ``detail free'' in the sense that no specific 
knowledge of the distributions is needed in addition to 
the fact that bidder $N+1$ is stronger than the others as required. 
In particular, no additional information about the value of the atom is needed.
This is important since mis-calibrating by a small amount
an attempt to implement the optimal auction can lead to large losses in revenue.

We provide an interpretation of these results as possibly providing guidelines to a seller
on how to strategically ``populate'' auctions 
with a single bidder even when only 
weaker bidders are
available. 
\blfootnote{$^{*}$Both authors are grateful to \href{https://www.eief.it/eief/}{EIEF} in Rome for hospitality and support. }
\end{abstract}

\section{Introduction}\label{Section: Introduction}
Since \citet{Vickrey:61} two benchmark auction formats have dominated a vast literature on auctions\footnote{We return
to a limited attempt to situate our contribution in this landscape in Section \ref{Section: Literature} below.}, the second price and the
first price auction (henceforth SA and FA respectively). This is so despite the fact that 
since \citet{Myerson:81} we know that neither format is optimal except in
very special circumstances.\footnote{See also \citet{Bulow-Roberts:89}.}

We consider a ``tournament'' auction (henceforth TA) in which $N$ $\geq$ $2$ bidders bid for the right to compete with bidder $N+1$. The highest
of the $N$ bidders competes in a second-price contest with $N+1$ for a single indivisible good. 
All bids are sealed, and the first-stage bids of the $N$
bidders are binding in the second-stage competition with $N+1$.

We focus on the case of a bidder $N+1$ that is sufficiently stronger than the first $N$, just as we did 
in \citet{Anderlini-Kim-EL:24}.
There, among other things, we studied the effects of a large $N$. Here we work with a fixed $N$, but
with a distribution of $N+1$'s values that approaches an atom above the upper bound of the distribution of values of the
first $N$ bidders. It turns out that our TA is able to approximate the \citet{Myerson:81} optimal auction (henceforth OA)
in a surprisingly informationally parsimonious (or ``detail free'') way.

We argue that our results can be interpreted as a way to give advice to a seller facing a single buyer. This will consist 
of guidelines on how to ``populate'' the auction with competing sellers even if they are all evidently weaker than
the original single buyer.

\section{A Shot in the Dark}\label{Section: A Shot in the Dark}
Many of the desirable properties of our TA when the distribution of $N+1$'s values becomes more and more concentrated
on an atom above the distribution of values of the first $N$ bidders can be captured by looking at an extreme case. This lacks the
``smoothness'' required for general comparisons in which we will be interested below but is intuitively instructive.

The first $N$ bidders 
are symmetric and have independent 
private values $v_i \overset{\mathrm{iid}}{\sim} F$ with support $[0,\bar{v}]$. 
The distribution $F$ is absolutely continuous with continuous density $f(v)$ $>$ $0$ $\forall v \in [0,\bar{v}]$.\footnote{Absolute continuity 
is not needed for our example in this section. We include it here simply because these
are our maintained assumptions on the distribution of the $v_i$s in the general case we consider below.}

The value $w$ of $N+1$
is drawn independently from the $v_i$s and has a discrete distribution 
given by
\begin{eqnarray}\label{eqn: bid zero and k in discrete example} 
    w = \begin{cases} k {\rm \; with \; probability \;} p \\
    0 {\rm \; with \; probability \;} 1-p 
    \end{cases}
\end{eqnarray}
with $k>\bar{v}$ and $p \in (0,1)$.

Ties naturally arise with positive probability in this example because of the discrete distribution of $w$. 
We assume that ties in the first stage are resolved randomly with a positive probability of winning for 
each tied bidder. A tie in the second stage instead is always resolved in favor of $N+1$.\footnote{This is obviously
special and greatly simplifies the analysis of this example. 
No such assumption of asymmetric resolution of ties will be needed in the general case.}

Denote by $b_i(\cdot)$ the bidding function of bidder $i=1, \ldots, N+1$. The following constitutes an equilibrium
for the TA in the example at hand. The proof can be found in the Appendix.
\begin{proposition}[Discrete Distribution]\label{prn: equilibrium in p and 1-p}
Suppose that $pk$ $>$ $\bar{v}$. Then
the following is an equilibrium\footnote{\rm The equilibrium is essentially
unique. For the first $N$ bidders, the value of $b_i(0)$ could be set at any level in $[0,k]$, 
but any non-zero bid would involve
playing a weakly dominated strategy. It also would not change the 
outcome of the TA, and hence it would not change the
revenue.}
for the TA in the example at hand.
 
Bidder $N+1$ bids his value, so that he sets $b_{N+1}$ $=$ $w$ for
every $w$ $\in$ $\{0,k\}$. For every $i$ $=$ $1, \ldots, N$ we have 
\begin{eqnarray}
   b_i(v_i) = \begin{cases} k {\rm \quad if \quad } v_i \in (0,\bar{v}] \\
    0 {\rm \quad if \quad} v_i = 0 
    \end{cases}
\end{eqnarray}
The expected revenue generated by this equilibrium is $p k$. It follows that as $p$ approaches $1$, the TA
is approximately optimal in terms of expected revenue.
\end{proposition}

A key property of the equilibrium in Proposition \ref{prn: equilibrium in p and 1-p} is the revenue it generates, 
approximating that of the OA as $p$ approaches $1$. The fact that $N+1$
bids $k$ when his value is $k$ is a straightforward consequence of the fact that the second stage is a second price contest and hence
bidding one's value is a weakly dominant strategy. This would be true in a standard second price auction as well of course.

If, with the same values, we opted for a standard SA instead of the TA, 
with probability $p$ the value of $k$ would be revealed ex-post
via $N+1$'s bid. However
the expected revenue would equal the expected value of first order statistic of 
the first $N$ values and hence would be bounded above by $\bar{v}$.
The value of $k$ would be revealed, but ``too late'' to be captured by the auctioneer's revenue. 

Of course, if the auction design could incorporate information about the actual value of $k$, then capturing the expected revenue as in the
TA would be straightforward. The auctioneer could set a {\it reserve price} of $r$ for $N+1$ ``close'' to (below) 
$k$, and thus obtain an expected revenue
bounded below by $p  r$. 

The difference between the TA we described and the SA with a reserve price lies in the information that is required
in the auction design. The TA we described does {\it not} require any information about the value of $k$ to generate revenue $p k$. 
While it  clearly requires
the knowledge of the identity of the strong bidder (namely $N+1$), it
is detail free in this crucial dimension. 

Setting
the reserve price in the standard second price auction in the absence of precise 
information about $k$ instead is very much a ``shot in the dark.'' 
Any distance between $r$ and $k$ ``from below'' is correspondingly costly in terms of expected revenue. 
Setting an $r$ above
$k$ decreases expected revenue discontinuously to a level bounded above by $\bar{v}$. 
In addition, in this case the outcome also fails to be ex-post
efficient since $N+1$ will not win when his value is $k$. Overall surplus (the sum of payoffs) also drops 
discontinuously once $r$ exceeds $k$.

The purpose of this paper is to show that the ability of the TA format to approximate 
the OA expected revenue generalizes considerably while 
retaining the informational parsimony showcased in the example, which the OA in general does not posses.

\section{Outline}\label{Section: Outline}
The material in the rest of the paper is organized as follows. 
In Section \ref{Section: Literature} we attempt to place our contribution in the
context of the enormous extant literature. In Section \ref{Section: Model} we 
present the model in full detail. Section \ref{Section: Preliminary Results}
presents some preliminary results, guaranteeing existence and 
presenting a general ``overbidding'' feature of the TA that underlies our results.

In Section \ref{section: Approximate Optimality Results} 
we begin by
discussing
the trade-off between optimality and ex-post efficiency
for a SA in our strong bidder set-up and then we
present our main result.
In a nutshell, the TA approximates the OA in 
our set-up. Crucially, the TA does not use any information about
$k$ other than the fact that it is greater than $\bar{v}$. 

In Section \ref{Section: Miscalibration Results} we examine
how attempting to use a SA with a reserve price to approximate
the OA is an enterprise fraught with difficulties. Miscalibrating the reserve price 
relative to the atom the in the limit distribution of $N+1$'s values 
can backfire in several ways. Two are the equivalent of what we highlighted
in the discrete distribution example of Section \ref{Section: A Shot in the Dark}. 
An additional one arises because the way the reserve price approaches $k$ 
is ``mismatched'' with the way the distribution of values for $N+1$ converges to an atom on $k$.

Section \ref{Section: Auctioneer Intervention} examines a modification of our basic TA
design that may be appealing in its own right and that sheds light on a key assumption of the model.

In Section \ref{Section: Populating Auctions}, we turn the interpretation of our
main result, so to speak, on its head and we conclude the paper. We argue at some length that our results
can be interpreted as providing some compelling ``advice'' to a seller facing a single 
buyer and thus an extreme version of the shot in the dark problem we mentioned above.


To streamline, all proofs have been gathered in an Appendix. Any item (equations and so on)
with a numbering that is prefaced by an ``A'' is to be found in the Appendix.

\section{Relation to the Literature}\label{Section: Literature}
The work-horses of a truly vast literature on auctions that has 
developed since \citet{Vickrey:61} are the
first- and second-price auctions (FA and SA respectively).\footnote{\label{fn: vast lit}The enormous
edited collection presented in \citet{Klemperer:00} is a good way to get a sense of the 
sheer size of the extant literature. \citet{McAfee-McMillan:87} systematically lay out the state of the literature in the late 80s.} 

A classic paper by \citet{Myerson:81} presents an elegant solution to the problem of finding the optimal auction (OA) for selling an
indivisible object to a given set of potential buyers. In general optimal auctions need to take into account many details 
of the distributions of values and do not fall neatly under either of the two main formats. 
The information on the distribution of values may need to be incorporated in ``personalized'' reservation prices and ``priority levels'' 
and this makes the OA difficult to implement and not suited to many practical applications.\footnote{See \citet{Bulow-Roberts:89}.}
This has guided a vast literature to examine
variations
of the two basic formats and their effects on bidders and revenue.

Our TA displays desirable properties in the case of asymmetric bidders, with one bidder ($N+1$) being stronger than
the first $N$ $\geq$ $2$. \citet{Maskin-Riley:00} studied asymmetric auctions. They focus on the comparison of 
SA and FA from both the bidders' and the seller's point of view. Their insights are difficult to compare with our set up
because they study a a two-bidder
environment. Instead our set up only makes sense if there are three
or more bidders ($N\geq2$).

 A vigorous empirical and theoretical literature
on procurement auctions going back to at least \citet{Hendrix-Porter:88} and extending all the way to the more recent 
\citet{Athey-Levin-Seria:11} (among others) is concerned with the effects of asymmetric information across bidders. In a procurement
set-up it seems natural to conjecture that incumbents have a substantial informational advantage over newcomers. The strong
$N+1$ bidder assumption seems to broadly fit a procurement environment with a known strong insider.

In \citet{Anderlini-Kim-EL:24}\footnote{See also the working paper that it replaced,
\citet{Anderlini-Kim-arxiv:24}.} we show that with a strong $N+1$ {\it overbidding}\footnote{Bidding above one's value.} on the part
of the first $N$ bidders arises naturally in the TA, giving it an advantage in terms of expected revenue over the SA. 
Surprisingly, this remains true even in the limit as $N$ becomes large. 

Here $N$ is given and we are concerned with a strong $N+1$ bidder with a distribution of values that converges to an atom. We are able to
compare the TA with the OA in the limit and show that expected revenue in the TA converges to the OA revenue. Unlike
the OA our TA design is parsimonious in terms of the bidders' distributions of values. What is needed is the knowledge that $N+1$
is the stronger bidder, that he is indeed sufficiently stronger and that his distribution of values is 
concentrated in the sense of being to an 
atom.\footnote{If we pursue further the idea that the stronger bidder is an insider in a procurement set-up, it seems natural 
to conjecture that past experience would make
his distribution of values correspondingly concentrated. 
This observation is related to our discussion in section \ref{Section: Populating Auctions}.} The actual value that the atom assigns to the object for sale is {\it not} an
input of the design of the TA.
In this sense the TA is informationally parsimonious or ``detail free.''\footnote{\citet{Bergemann-Morris:13} investigate the robustness of the 
equilibrium outcomes of Bayesian games to their information structure. Their set up is extremely different from the one we 
analyze here, but our informal notion of informational parsimony (``detail free'') is of course related to the notion of robustness
in their work.}





\section{Model}\label{Section: Model}
\subsection{Baseline Model}\label{Subsection: Baseline}
A single indivisible object is for sale and 
there are $N+1 \geq 3$ bidders. The first $N$ compete for the right to enter
a second-price contest with $N+1$. The highest of the first $N$ bids $i^*$ wins and proceeds to the second stage.
The highest of $b_{i^*}$ and $b_{N+1}$ wins the object and pays a price equal to the bid of the other.

All bids are sealed and binding, and any ties (in the first or second stage) are resolved by a random draw that
gives positive probability of winning to all tied bids.

The values of the first $N$ bidders are exactly as in the discrete distribution example 
of Section \ref{Section: A Shot in the Dark}.
The second-stage bidder $N+1$ has value $w \sim G$ with support $[0,\bar{w}]$. 
The distribution $G$ is absolutely continuous with density $g(w)$ $>$ $0$ 
$\forall w \in [0,\bar{w}]$ that is 
continuously differentiable on $[0,\bar{w}]$. The value $w$ is drawn independently 
of the values $v_i$ of the first stage bidders. 
The second-stage bidder is ``stronger'' in the sense that 
$E[w] \geq \bar{v}$.

These assumptions will be maintained throughout.

\subsection{Convergence}\label{Subsection: Convergence}
Our main focus is the behavior of the baseline model when the distribution $G$ of $N+1$'s values converges to an atom
above $\bar{v}$. To do this, we define a sequence of random variables $\{w_{l}\}_{l=1}^{\infty}$ with distributions $G_{l}$
each satisfying the maintained assumptions of the baseline model. Moreover,
we will assume that 
\begin{eqnarray}\label{eqn: convergence in distribution}
    w_l \xrightarrow{d} k \; > \; \bar{v}
\end{eqnarray}
and\footnote{In the proof of Proposition
\ref{Proposition: Tournament Auction Revenue} it will
be apparent that 
(\ref{eqn: conditional mean converges to zero}) is 
equivalent to stipulating that $\lim_{l \to \infty}$ $E[w_l \, |
\, w_l \leq c]$ $=$ $0$ $\forall$ $c$ $\in$ $(0,k)$.}
\begin{eqnarray}\label{eqn: conditional mean converges to zero}
    \lim_{l \to \infty} 
    \frac{G_l(c_2) -G_l(c_1)}{G_l(c_2)}
    = 0 \quad {\rm whenever} \; 0< c_1 < c_2 < k
\end{eqnarray}

Condition (\ref{eqn: convergence in distribution}) is a standard one that stipulates that the sequence
$\{w_{l}\}_{l=1}^{\infty}$ converges in distribution to a degenerate random variable that assigns probability $1$ to
the value $k$. Condition (\ref{eqn: conditional mean converges to zero}) is less familiar. Intuitively, we require that along the sequence the probability
mass is drained away from low values sufficiently slowly. We return to the intuitive role of
this condition in Sections \ref{section: Approximate Optimality Results} and \ref{Section: Miscalibration Results}. 
Section \ref{Section: Auctioneer Intervention} discusses a modification of our TA design that would allow
(\ref{eqn: conditional mean converges to zero}) to be dispensed with entirely.
For the time being,
we notice that an appropriately ``smoothed out'' version of the discrete distribution 
we used in Section \ref{Section: A Shot in the Dark}
will fit the bill.

\section{Preliminary Results}\label{Section: Preliminary Results}


Since the second stage of the TA is a second price contest, it follows from standard arguments that 
it is a {\it weakly dominant} strategy for $N+1$ to set $b_{N+1}$ $=$ 
$w$.\footnote{\label{fn: truth is still weakly dominant when reserve price is set}For future reference,
note that weak dominance remains true when the auction is modified so that a {\it reserve price} is set for
$N+1$.}
Throughout, we will 
assume that in equilibrium this is indeed the case. In the description of any equilibrium, 
we will actually omit to specify $b_{N+1}(\cdot)$, it being understood that it is truthful as above.

We will also restrict attention throughout to equilibria that are symmetric in the sense that $b_i(\cdot)$ is the
same for every $i$ $=$ $1,\ldots,N$, and when this does not cause ambiguity, we will just denote this
common bidding function by $b(\cdot)$. Moreover, we will restrict attention to the case in which
$b(\cdot)$ is strictly increasing. Both these restrictions are without loss of generality. 
However, due to the novel format of the TA, the
existence of such a $b(\cdot)$ requires proof.

\begin{proposition}[Existence]\label{Proposition: Existence}
Under our maintained assumptions, there exists a $b(\cdot)$
that constitutes a symmetric equilibrium of the TA. Moreover, any such $b(\cdot)$
must be strictly increasing.
\end{proposition}

The fact that the first $N$ bidders in equilibrium bid above their respective values 
is a key feature of the TA. It underlies
the results in \citet{Anderlini-Kim-EL:24}, and the approximate optimality 
one we present below in Section
\ref{section: Approximate Optimality Results}. It appears with its proof 
as a Lemma (the only one) in the online appendix to \citet{Anderlini-Kim-EL:24}. We reproduce it here
purely for the sake of completeness.

\begin{proposition}[Overbidding]\label{Proposition: Overbidding}
Let $b(\cdot)$ be a strictly increasing function that 
constitutes an equilibrium for the TA under our maintained assumptions.
Then
\begin{eqnarray}\label{eqn: overbidding}
b(v) \; > \; v \qquad \forall \, v \in (0,\bar{v}]
\end{eqnarray}
\end{proposition}

Intuitively, the equilibrium overbidding of Proposition \ref{Proposition: Overbidding}, is the
result of the interplay between the two-stage structure of the TA and the second-price nature of the second-stage contest.

The only way for one of the first $N$ bidders with value $v_i$ to achieve a positive payoff is to {\it win} the first stage. This generates acute
competition among the $N$ weaker bidders. The potential downside of bidding above value in order to win the first stage is mitigated
by the nature of the second stage. Since this is a second price contest, with some probability the price paid will still be below $v_i$.
The balance of these two forces yields the optimal bid $b_i(v_i)$, which under the maintained assumptions of Subsection
\ref{Subsection: Baseline} will be above $v_i$.

\section{Approximate Optimality of the Tournament Auction}\label{section: Approximate Optimality Results}

Our results in this and the next section concern auctions with a sequence of distributions of values.
Throughout this and the next subsection we 
work with our maintained assumptions of Subsection \ref{Subsection: Baseline} and with a
sequence of $\{w_l\}_{l=1}^{\infty}$ satisfying the convergence conditions (\ref{eqn: convergence in distribution}) 
and (\ref{eqn: conditional mean converges to zero}). 

Some new notation is needed. We let $R_l^{TA}$, $R_l^{OA}$ and $R_l^{SA}$ be the 
expected revenue of the Tournament, Optimal and Second Price auctions respectively along
the sequence $\{w_l\}_{l=1}^{\infty}$. Similarly  $S_l^{TA}$, $S_l^{OA}$ and $S_l^{SA}$ will be
the expected total surplus (expected sum of payoffs) generated by the three auction
formats along the sequence. Finally, $v_{q:N}$ is the $q$-th order statistic of the random variables
$v_1, \ldots, v_N$. 

An OA need not be efficient.\footnote{Efficient here means maximizing
the expected total sum of payoffs. Ex-post efficient has the usual meaning
of selling to the highest valuer with probability one.}
It may well be that that to maximize expected revenue it may be optimal to set a reserve price that induces the good not to be sold
in some cases, thus creating an inefficiency. This is a pervasive phenomenon, but as \citet{Myerson:81} 
points out it is especially apparent and
intuitive in the case of a single bidder.\footnote{We will return to single bidders below in Section
\ref{Section: Populating Auctions}.}

The trade-off between revenue and efficiency in our context is well exemplified by the SA. Clearly,
the expected revenue of the SA is bounded above by $\bar{v}$ $<$ $k$. On the other hand
the SA is ex-post efficient, and hence overall efficient. Given that $\{G_l\}_{l=1}^{\infty}$
converges in distribution to to $k$ as in  (\ref{eqn: convergence in distribution}), this easily implies that
$S_l^{SA}$ converges to $k$.

It is interesting to notice that while the trade off between revenue and efficiency is present before we reach
the limit, our set up implies that there the trade off vanishes the limit. 
Intuitively if the limit expected revenue is $k$, then it must be that the
$N+1$ bidder is the one that obtains the object with probability closer and closer to $1$. This is 
simply because $k$ $>$ $\bar{v}$, and $N+1$ is the only bidder who can possibly end up
paying a price near $k$ in any equilibrium. 

The intuition we just proposed concerning the vanishingly small trade off between optimality and efficiency turns out to be correct,
as our next proposition shows.
\begin{proposition}[No Limit Inefficiency]\label{Proposition: No Limit Inefficiency}
Consider a sequence of auction mechanisms that generate a sequence of equilibrium expected revenues $\tilde{R}_l$
such that $\lim_{l \rightarrow \infty} \tilde{R}_l$ $=$ $k$, and let $\tilde{S}_l$ be the corresponding sequence of expected
surpluses. Then
\begin{eqnarray}\label{eqn: limit tilde suplus equals k}
\lim_{l \rightarrow \infty} \tilde{S}_l \; = \; k
\end{eqnarray} 
\end{proposition}

Somewhat unsurprisingly, the OA yields revenue $k$ in the limit.
\begin{proposition}[Optimal Auction Revenue]\label{Proposition: Optimal Auction Revenue}
\begin{eqnarray}\label{eqn: limit optimal revenue equals k}
\lim_{l \rightarrow \infty} R^{OA}_l \; = \; k
\end{eqnarray}
\end{proposition}

Combining Propositions \ref{Proposition: No Limit Inefficiency} and \ref{Proposition: Optimal Auction Revenue}
we immediately know that the OA is in fact efficient in the limit. Formally
\begin{eqnarray}\label{eqn: limit revenue of OA equals k}
\lim_{l \rightarrow \infty} {S}^{OA}_l \; = \; k
\end{eqnarray}

The next proposition embodies our main result.
It relies on the systematic overbidding of the first $N$ bidders
identified in Proposition \ref{Proposition: Overbidding} above, and it generalizes what we saw in the 
discrete distribution example of Proposition \ref{prn: equilibrium in p and 1-p}. 

\begin{proposition}[Tournament Auction Revenue]\label{Proposition: Tournament Auction Revenue}
\begin{eqnarray}\label{eqn: limit tournament revenue equals k}
\lim_{l \rightarrow \infty} R^{TA}_l \; = \; k
\end{eqnarray}
\end{proposition}

The TA also exhibits a lack of limit trade off between revenue and efficiency since it approximates the OA.
Proposition \ref{Proposition: Tournament Auction Revenue} generalizes Proposition \ref{prn: equilibrium in p and 1-p}
to the smooth case that allows a proper comparison with the OA of \citet{Myerson:81}. 

The discrete example of Section \ref{Section: A Shot in the Dark} helps to understand intuitively the role
that (\ref{eqn: conditional mean converges to zero}) plays in the argument behind Proposition 
\ref{Proposition: Tournament Auction Revenue}. Condition (\ref{eqn: conditional mean converges to zero}) ensures
that ``enough'' probability is left on ``low values'' of $w$ as we approach the limit atom on $k$ in a way that parallels 
the probability mass $1-p$ left on $w=0$ in the discrete example of Section \ref{Section: A Shot in the Dark}. This 
in turn constitutes enough of an incentive (the likelihood of a low second stage sale price) 
for the first $N$ players to bid sufficiently near $k$ as we approach the limit (to bid $k$
in the discrete example) to reach the conclusion in of Proposition \ref{Proposition: Tournament Auction Revenue}.

As we emphasized before,
in our view the key feature of the TA is that it's design does {\it not} utilize any information about the value of $k$. The
``shot in the dark'' problem that we found in Section \ref{Section: A Shot in the Dark} is solved by the TA
in the general case.

\section{Miscalibration}\label{Section: Miscalibration Results}

\citet{Myerson:81} shows that the OA for a seller facing a single buyer is without loss of generality
characterized by a single reserve price $r$. The seller announces $r$, and then the buyer
either buys the object at $r$ or the object remains unsold. The reserve price depends on the entire
distribution of values for the single buyer.

Our set up is in some sense close to the single buyer case, especially when we
are close to the limit as $l$ becomes large and $N+1$'s value
approaches the atom on $k$. A natural question to ask is then the following. 
Can the OA be approximated by picking an 
appropriate sequence of reserve prices in a second price
auction setting? This can in fact be done, but it is subject to a 
``shot in the dark'' miscalibration problem that is even more pervasive
than the one that we identified in the discrete distribution example 
of Section \ref{Section: A Shot in the Dark}.


We are interested in the general version of the miscalibration question we examined 
in Section \ref{Section: A Shot in the Dark}. To 
do this we consider the sequence of distributions $G_l$ as in 
Subsection \ref{Subsection: Convergence} satisfying
the convergence properties set out in (\ref{eqn: convergence in distribution}) 
and (\ref{eqn: conditional mean converges to zero}),
together with an sequence of reserve prices $\{r_l\}_{l=1}^{\infty}$. 
The objective is to examine the expected revenue and surplus of
the sequence of second price auctions with values $G_l$ and reserve prices $r_l$ that apply to
bidder $N+1$. 
These will be denoted by
$R^{SA}_l(r_l)$ and $S^{SA}_l(r_l)$ respectively.

Our first result says that when $G_l$ is close to the limit it is always possible to chose
$r_l$ so that the TA is almost optimal in terms of revenue.

\begin{proposition}[Approximating Sequence]\label{prn: Approximating Sequence}
There exists a sequence $\{r_l\}_{l=1}^\infty$ such that 
\begin{eqnarray}\label{eqn: limit k is achieved}
\lim_{l \to \infty} R_l^{SA}(r_l) = k
\end{eqnarray}
Any such sequence must satisfy $\lim_{l \to \infty} r_l = k$.
\end{proposition}

The perils of undershooting $k$ or overshooting it are the exact analogues of what the
discrete example of Section \ref{Section: A Shot in the Dark} suggests. Undershooting
is costly in terms of expected revenue, and the magnitude of the cost is
continuously related to the amount of undershooting. Efficiency is guaranteed in the limit.

\begin{proposition}[Undershooting]\label{prn: Undershooting}
Suppose that $\lim_{l \to \infty} r_l = \bar{r}$ $\in$ $(\bar{v},k)$. Then

\begin{eqnarray}\label{eqn: undeshooting revenue below and surplus ok}
    \lim_{l \to \infty} R_l^{SA}(r_l) = \bar{r} < k
\qquad
{\rm and}
\quad
    \lim_{l \to \infty} S_l^{SA}(r_l) = k
\end{eqnarray}
\end{proposition}

On the other hand, overshooting is much more perilous than undershooting. Any amount 
of overshooting yields a discontinuously lower expected revenue and surplus. Bidder $N+1$
is just not willing to buy at the reserve price. 

\begin{proposition}[Overshooting]\label{prn: Overshooting}
Suppose that $\lim_{l \to \infty} r_l = \bar{r}$ $>k$. Then
\begin{eqnarray}\label{eqn: overshooting revenue way below}
    \lim_{l \to \infty} R_l^{SA}(r_l) = E[v_{2:N}] < E[v_{1:N}] = \lim_{l \to \infty} R_l^{SA} < \bar{v} <  k
\end{eqnarray}
and
\begin{eqnarray}\label{eqn: overshooting surplus below}
    \lim_{l \to \infty} S_l^{SA}(r_l) = E[v_{1:N}]  < \bar{v} < k
\end{eqnarray}

\end{proposition}

From Proposition \ref{prn: Approximating Sequence} we see that while (\ref{eqn: limit k is achieved})
implies that $\lim_{l \to \infty} r_l = k$, the reverse may not be true. This is in fact the case.
Not all sequences $\{r_l\}$ that converge to $k$ suffice to approximate the OA expected revenue as in
Proposition \ref{prn: Approximating Sequence}. Convergence of $\{r_l\}$ to $k$ is necessary but not sufficient.

Given Propositions \ref{prn: Undershooting} and 
\ref{prn: Overshooting}, this is particularly surprising when the sequence $\{r_l\}$ approaches $k$ from
{\it below} since a limit reserve price below $k$ is not discontinuously costly. This is the focus of our next proposition.

\begin{proposition}[Convergence From Below]\label{prn: Convergence from Below}
Fix a $p$ $\in[0,1]$. For any such $p$ there exists a sequence $\{G_l\}$ converging to 
$k$ as in (\ref{eqn: convergence in distribution}) and
(\ref{eqn: conditional mean converges to zero}) and a sequence of reservation prices
$\{r_l\}_{l=1}^{\infty}$ with $r_l \rightarrow k$ and $r_l$ $<$ $k$ $\forall$ $l$ such that
\begin{eqnarray}
    \lim_{l \to \infty} R_l^{SA}(r_l) = p E[v_{2:N}] + (1-p) k
\end{eqnarray}
\begin{eqnarray}
    \lim_{l \to \infty} S_l^{SA}(r_l) = p E[v_{1:N}] + (1-p)k
\end{eqnarray}
\end{proposition}

Intuitively condition (\ref{eqn: conditional mean converges to zero}) plays a key role in the argument.
As $G_l$ approaches the atom $k$, it could be that enough probability is left ``near zero''
and that $r_l$ approaches $k$ fast enough so that the combined effect of this speed mismatch
is almost the same as a reserve price set above $k$.

There is a sense in which Proposition \ref{prn: Convergence from Below} indicates that 
``a lot'' of information about the value of $k$ and $G_l$ may need to be taken into account
for the SA with a reserve price auction design to be failsafe. This is not the case for the TA.

\section{Detail Free ``Auctioneer Intervention''}\label{Section: Auctioneer Intervention}
As we anticipated, we return to the role and interpretation of our assumption 
that as $w_l$ converges to an atom $k$, the rest of the distribution should be drained away from low values
sufficiently slowly --- namely condition (\ref{eqn: conditional mean converges to zero}) in Subsection
\ref{Subsection: Convergence}.  
We do this informally since the argument that substantiates our claim is a minor modification of the proof of 
Proposition \ref{Proposition: Tournament Auction Revenue}, our main result. Our discrete example
of Section \ref{Section: A Shot in the Dark} provides a strong intuition as well.

Suppose that we have a sequence of auctions with a strong bidder satisfying our baseline assumptions
of Subsection \ref{Subsection: Baseline} and with $\{w_l\}_{l=1}^{\infty}$ converging to an atom $k$, satisfying
(\ref{eqn: convergence in distribution}) (convergence in distribution) but not necessarily (\ref{eqn: conditional mean converges to zero})
(slow drain from low values). 

We now let the auctioneer ``intervene'' in the TA design in an essentially detail free way. By this we mean in a way that only uses 
``low grade information'' about $k$. In particular, if the auctioneer knows that $k$ $>$ $\bar{v}$ then he will be able to 
pick a $p$ $\in$ $(0,1)$ such that $p k$ $>$ $\bar{v}$.\footnote{It is interesting to notice that, trivially, the ``tightness'' of the condition
$p k$ $>$ $\bar{v}$ goes down when $k$ is larger and/or $\bar{v}$ is smaller. It is easier to satisfy as $N+1$ becomes more and more
stronger than the first $N$ bidders.}
Then the auctioneer can modify the TA design by announcing in advance
that after the sealed bids have been submitted he will use a randomization device that will change $N+1$'s bid to be {\it equal to zero}
with probability $1-p$. After the draw has been executed the rules of the TA are exactly as in our analysis so far --- the winner of the first $N$
bidders enters a second price context with the (modified if appropriate) bid of $N+1$. 

The limit (as $l$ $\rightarrow$ $\infty$) expected revenue of this modified TA can be easily seen to be equal to $p k$. Thus, as the
distribution of values of $N+1$ becomes more concentrated on $k$, the auctioneer can capture a sizable fraction of the surplus $k$. 
Condition (\ref{eqn: conditional mean converges to zero}) can be dispensed with by modifying the TA design with the randomized
auctioneer intervention we have described.

\section{Populating Auctions with a Strong Bidder}\label{Section: Populating Auctions}
We conclude the paper describing a further possible extension of the model that we believe
also delivers a key insight into what drives our results at a basic ``information structure'' level.

A key observation in \citet{Myerson:81} characterizes the OA in the 
case of a single bidder. In this case the seller should offer to sell at a given 
price and keep the object if the buyer is not willing to buy at that price. The computation
that yields the reserve price of the optimal single buyer uses a lot of information
about the distribution of the buyer's value. It is intuitively not unlike the computation that
a monopolist carries out, except that the marginal revenue is a probabilistic object in the auction case.

For the reasons we discussed intuitively in Section \ref{Section: Miscalibration Results}
if the distribution of values is very concentrated (nearing an atom $k$) the possibility of
miscalibrating the reserve price yields potentially discontinuously large revenue losses for
the seller. Information about $k$ is very valuable to the seller.

Of course, there is a potential pitfall in our reasoning above. If the seller does not
have good information about $k$, then why is the distribution of his values concentrated around
$k$? Without going into any formal details (they are standard), one can resolve the issue using a device
familiar from the literature on auctions with affiliated values. We can imagine Nature first drawing $k$
and then (independently) the actual value of the buyer from a distribution narrowly concentrated around $k$.

The two stage draw we have described is of course without consequence if the auction remains
one with one buyer and the seller remains uninformed about $k$. 
However, we think our main result above (Proposition \ref{Proposition: Tournament Auction Revenue})
suggests a way in which the seller may attempt to ameliorate his position in this case. Suppose
that the seller who does not know $k$, knows instead that there are {\it potential} bidders that,
while being {\it weaker} than the original one, have accurate information about it --- they
know $k$ in the sense that they observe the outcome of the first draw that determines $k$.\footnote{The 
discrete example of Section \ref{Section: A Shot in the Dark} tells us that the potential weak bidders
need not know $k$ with certainty. They could just receive a sufficiently precise signal (binary) about the draw that
determines $k$. If the second stage draw is degenerate this reproduces that exact information structure of
Section \ref{Section: A Shot in the Dark}, but a
moderately more general statement holds.}
We mentioned
procurement with a strong insider in Section \ref{Section: Literature} discussing related literature. This set up
seems to fit well the scenario we are informally describing here.

The potential bidders may simply not be aware of the opportunity to bid, or may be not sufficiently incentivized
to participate because they know themselves to be much weaker than the insider. Proposition 
\ref{Proposition: Tournament Auction Revenue} says that somewhat surprisingly the seller may stand to 
gain very substantially by somehow attracting the weaker bidders to participate. In fact 
Proposition 
\ref{Proposition: Tournament Auction Revenue} says that in the limit the seller will be able to extract
the entire surplus $k$ from the original single buyer.\footnote{In a very different 
context \citet{Milgrom-Weber:82} find that the equilibrium features of the auction they analyze 
may provide support for the fact that the seller may want attempt to change the informational structure of the 
game by providing ``expert advice'' to buyers.}

We believe this observation about enticing weaker but well informed participants
to be interesting in its own right. However, this also sheds a further light on a key insight provided by Proposition 
\ref{Proposition: Tournament Auction Revenue}. What drives the near optimality of the TA is that,
under the right conditions, it allows
the seller to leverage to {\it his} advantage the information of the weaker bidders {\it even if he does not have the information
himself}. In a nutshell, this is what drives the informationally parsimonious nature of the 
approximately optimal TA format that we have investigated in this paper. 

\newpage

\fontsize{12}{13.89}\selectfont
\bibliographystyle{ectranew}
\bibliography{large}

\newpage{}

\begin{appendix}

\section{The Proof of Proposition \ref{prn: equilibrium in p and 1-p}}\label{section app: proof of prn equilibrium in p and 1-p}
Assume that bidders $-i$ bid according to (\ref{eqn: bid zero and k in discrete example}) and consider bidder $i$ $=$ $1, \ldots , N$
with value $v_i$ $=$ $0$. If he bids $0$ as the proposed equilibrium requires, his expected payoff is $0$. If he bids $b_i$ $>$ $0$ and he wins
clearly his payoff cannot be greater than zero since the price cannot be negative. If he bids $b_i$ $>$ $0$ and he does not win, then his payoff is
zero. Hence he has no incentive to deviate from the proposed equilibrium when $v_i$ $=$ $0$.

Now consider bidder $i$ with value $v_i$ $>$ $0$. If he bids $k$ as the proposed equilibrium requires, he wins the first stage with positive probability
as the tie with the other first stage bidders is resolved in his favor with positive probability, say $m$ $\in$ $(0,1)$. He then wins the object 
in the second stage if and {\it only if} the bid of $N+1$ is $0$. In this case
he gets the object and pays 
a price of $0$. So, by bidding $k$ as required $i$ has an expected payoff of $(1-p) m v_i$. If instead $i$ bids $b_i$ $<$ $k$, he loses the first stage 
contest to the other $-i$ bidders. Hence his expected payoff in this case is $0$. If he bids $b$ $>$ $k$,
he wins the first and second stage for sure and pays a price equal to the bid of $N+1$. 
Therefore in this case his expected payoff is $v - pk$ $\leq$ $\bar{v} - pk$ $<$ $0$.
This is clearly enough to prove the proposition.\endproof

\section{The Proof of Proposition \ref{Proposition: Existence}}\label{section app: proof of prn Existence}
The following are standard, but we repeat the statements here for the sake of completeness since they are used in the argument that follows.

\begin{lemma}[Schauder's Fixed Point Theorem] Suppose $C$ is a closed convex nonempty subset 
of a normed linear space and $D$ is a relatively compact subset of $C$. Let $F:C \rightarrow D$ be continuous. Then, $F$ contains a fixed point.
\end{lemma}

\begin{lemma}[Arzel\`a-Ascoli Theorem] A family of continuous real-valued
functions on $[a,b]$ with $-\infty<a<b<\infty$ is relatively compact under the sup-norm if and only if it is uniformly bounded and uniformly equicontinuous.
\end{lemma}

The actual proof of Proposition \ref{Proposition: Existence} is divided in 6 steps. Steps 1-5 establish existence, and Step 6 proves that any $b(\cdot)$ inducing an equilibrium must be strictly increasing.

\begin{step}\label{step: Phi}
We define the function $\Phi$ and establish some of its key properties for use in future steps.
\end{step}
Define the function $\Phi:[0,\bar{w}] \rightarrow [0,E(w)]$ as below:
\begin{eqnarray}
    \Phi(b) = E[w|w \leq b].
\end{eqnarray}
It is immediately clear that $\Phi$ is differentiable on $(0,\bar{w}]$ with
\begin{eqnarray}
    \Phi'(b) = \frac{g(b)}{G(b)}[b-\Phi(b)]>0 \quad \forall b \in (0,\bar{w}].
\end{eqnarray}
Furthermore, we can show that differentiability also holds at $0$ with
\begin{eqnarray}
\begin{split}
    \Phi'(0) & = \lim_{b \downarrow 0} \frac{E[w|w \leq b]}{b} = \lim_{b \downarrow 0} \frac{\int_0^b \xi g(\xi) \,d\xi}{bG(b)} \\
    &= \lim_{b \downarrow 0} \frac{bg(b)}{bg(b) + G(b)} = \frac{1}{1 + \lim_{b \downarrow 0} \frac{G(b)}{bg(b)}} = \frac{1}{2}
\end{split}
\end{eqnarray}
by applying L'Hopital's rule twice along the way.

We can then also check that $\Phi'$ is continuous on $[0,\bar{w}]$. This is immediate on $(0,\bar{w}]$. Then, also at 0, we can see that
\begin{eqnarray}
    \lim_{b \downarrow 0} \Phi'(b) = \lim_{b \downarrow 0} \frac{g(b)b}{G(b)}\Big[1- \frac{\Phi(b)}{b}\Big] = 1 - \Phi'(0) = \Phi'(0).
\end{eqnarray}

Hence we may conclude that $\Phi$ is continuously differentiable wit ha strictly positive derivative on $[0,\bar{w}]$. It then immediately follows that $\Phi$ is invertible with $\Phi^{-1}:[0,E(w)] \rightarrow [0,\bar{w}]$. Also, it follows that $\Phi^{-1}$ is continuously differentiable with a strictly positive derivative on $[0,E(w)]$. In particular, we note for future reference that $(\Phi^{-1})'(0) = 1/\Phi'(0) = 2$.

Finally, we note that
\begin{eqnarray}
    \Phi(b) < b \quad \forall b \in (0,\bar{w}]
\end{eqnarray}
\begin{eqnarray}
    \Phi^{-1}(\xi) > \xi \quad \forall \xi \in (0,E(w)].
\end{eqnarray}

\begin{step}
We now define an ODE and some related functions that will be useful for future purposes.
\end{step}

The ODE is defined as below:
\begin{eqnarray}
\begin{array}{c}\label{equation: ODE}
    b'(v) = H(b(v),v) \\
    H(b,v) \equiv (N-1) \frac{1}{b-v} \frac{f(v)}{F(v)} \frac{G(b)}{g(b)} \{v - \Phi(b)\} \quad \forall \; v \in (0,\bar{v}], \; b \in (v, \Phi^{-1}(v)).
\end{array}
\end{eqnarray}

Notice that any $b:(0,\bar{v}] \rightarrow \mathbb{R}_+$ satisfying the above ODE would immediately have $\lim_{v \downarrow 0} b(v) = 0$. Hence we refer to $b:[0,\bar{v}] \rightarrow \mathbb{R}_+$ as a solution to the ODE above if it satisfies (\ref{equation: ODE}) on $(0,\bar{v}]$ and $b(0)=0$. Notice that any solution to the ODE above must be strictly increasing on $[0,\bar{v}]$ since $H(b,v)>0$ on the defined region.

In order to establish our fixed point argument, it is convenient to work on the space $(b/v,v)$ rather than $(b,v)$. Hence we define the function
\begin{eqnarray}
    K(\beta,v) = H(\beta v,v) \quad \forall \; v\in (0,\bar{v}], \; \beta \in (1,\Phi^{-1}(v)/v).
\end{eqnarray}
Notice then that $K$ is partially differentiable with respect to $\beta$ and that
\begin{eqnarray}\label{equation: K beta partial derivative}
    \frac{\partial K(\beta,v)}{\partial \beta} = (N-1)f(v) \frac{v}{F(v)} \bigg[-1-\frac{[G(\beta v) - \frac{1}{v}\int_0^{\beta v} \xi g(\xi) \,d\xi][\frac{g(\beta v)}{v} + (\beta-1)g'(\beta v)]}{(\beta-1)^2 g(\beta v)^2}\bigg]
\end{eqnarray}
for all $v \in (0,\bar{v}], \beta \in (1, \Phi^{-1}(v)/v)$.

\begin{step}\label{step: local existence}
    There exists $\hat{\delta} \in (0,\bar{v}]$ and $b: [0,\hat{\delta}] \rightarrow \mathbb{R}_+$ such that $b$ satisfies (\ref{equation: ODE}) on $(0,\hat{\delta}]$ and $b(0)=0$.
\end{step}
This step is proved in 9 sub-steps, labelled from 3-a to 3-i.

{\bf Step \ref{step: local existence}-a}: There exists $\delta_1 \in (0,\bar{v}]$ such that for every $v \in (0,\delta_1]$, the function $K(\beta,v)$ is strictly decreasing in $\beta$ on $(1,\Phi^{-1}(v)/v)$.

Consider (\ref{equation: K beta partial derivative}). For all $v \in (0,\bar{v}], \beta \in (1,\Phi^{-1}(v)/v)$ we have that
\begin{eqnarray}
    G(\beta v) - \frac{1}{v} \int_0^{\beta v} \xi g(\xi) \,d\xi = [v-\Phi(\beta v)]\frac{G(\beta v)}{v} > [v - \Phi(\Phi^{-1}(v))]\frac{G(\beta v)}{v} = 0.
\end{eqnarray}

Furthermore, let us then define the region $R_\delta \equiv \{(\beta,v): v \in(0,\delta], \beta \in (1,\Phi^{-1}(v)/v)\}$ for any $\delta \in (0,\bar{v}]$. Then, since $g'$ is continuous and $\lim_{v \downarrow 0} \Phi^{-1}(v)/v = 2$ (see step \ref{step: Phi}), we have that
\begin{eqnarray}
    \lim_{\delta \downarrow 0} \inf_{(\beta,v) \in R_\delta} (\beta-1) g'(\beta v) \geq -|g'(0)|.
\end{eqnarray}
Also, we can see from the continuity of $g$ that
\begin{eqnarray}
    \lim_{\delta \downarrow 0} \inf_{(\beta,v) \in R_\delta} \frac{g(\beta v)}{v} = + \infty.
\end{eqnarray}
Thus, we can find $\delta_1 \in (0,\bar{v}]$ such that
\begin{eqnarray}
    \frac{g(\beta v)}{v} + (\beta - 1 )g'(\beta v) \geq 0
\end{eqnarray}
for all $(\beta,v) \in R_{\delta_1}$. Then, examining (\ref{equation: K beta partial derivative}), it is clear that we have
\begin{eqnarray}
    \frac{\partial K(\beta,v)}{\partial \beta} <0
\end{eqnarray}
for all $(\beta,v) \in R_{\delta_1}$, which clearly establishes the claim of this sub-step.

{\bf Step \ref{step: local existence}-b}: We now construct the bounded space of functions on which we will later establish our fixed point existence argument.

Firstly, observe that for every $v \in (0,\delta_1]$,
\begin{eqnarray}
    \lim_{\beta \downarrow 1} K(\beta,v) = +\infty
\end{eqnarray}
\begin{eqnarray}
    \lim_{\beta \uparrow \Phi^{-1}(v)/v} K(\beta,v) = 0.
\end{eqnarray}

Then, by step 3-a above, it is clear that for every $v \in (0,\delta_1]$, there exists a unique $\beta^*(v) \in (1,\Phi^{-1}(v)/v)$ such that $K(\beta^*(v),v) = 2N/(N+1)$. Furthermore, we define $\beta^*(0) = 2N/(N+1)$ so that $\beta^*[0,\delta_1] \rightarrow \mathbb{R}$ is a well-defined function.

We then prove $\beta^*$ is continuous. Continuity at $(0,\delta_1]$ is obvious, so we focus on continuity at $0$. Suppose, without loss of generality, that there is some $\epsilon>0$ and $\{v_n\}_{n=1}^\infty \subseteq (0,\delta_1]$ such that $v_n \downarrow 0$ and $\beta^*(v_n) > \beta^*(0) + \epsilon$ for all $n$. Then, due to Step 3-a, we have that
\begin{eqnarray}
    K(\beta^*(0)+\epsilon,v_n) > K(\beta^*(v_n),v_n) = \frac{2N}{N+1}
\end{eqnarray}
and so
\begin{eqnarray}
    \lim_{n \to \infty} K\Big(\frac{2N}{N+1}+\epsilon,v_n\Big) \geq \frac{2N}{N+1}.
\end{eqnarray}
However, notice that for all $\beta \in (1,2)$, we have
\begin{eqnarray}
\begin{array}{c}
    \lim_{v \downarrow 0} K(\beta,v) = (N-1) \frac{\beta}{\beta-1} \lim_{v \downarrow 0} \Big[\frac{f(v)v}{F(v)} \frac{G(\beta v)}{\beta v g(\beta v)} \Big(1-\frac{\Phi(\beta v)}{v}\Big) \Big] \\
    = (N-1) \frac{\beta}{\beta -1} (1-\Phi'(0)\beta) \\
     = (N-1) \frac{\beta}{\beta-1} (1-\frac{1}{2}\beta),
\end{array}
\end{eqnarray}
where the limit was computed based on results from step \ref{step: Phi}. Notice then that we can show
\begin{eqnarray}
    \lim_{v \downarrow 0} K\Big(\frac{2N}{N+1}+\epsilon,v\Big)<\frac{2N}{N+1},
\end{eqnarray}
which is clearly a contradiction. Hence we have established the continuity of $\beta^*$.

Given that $1<\beta^*(0) = 2N/(N+1) < 2  = \lim_{v \downarrow 0} \Phi^{-1}(v)/v$, we can then choose $\hat{\delta} \in (0,\bar{v}]$ small enough such that
\begin{eqnarray}
    1 < \min_{v \in [0,\hat{\delta}]} \beta^*(v) \leq \max_{v \in [0,\hat{\delta}]} \beta^*(v) < \inf_{v \in (0,\hat{\delta}]} \frac{\Phi^{-1}(v)}{v}.
\end{eqnarray}

We now define the functions
\begin{eqnarray}
    \overline{\beta}(v) \equiv \Big(\max_{\xi \in [0,v]} \beta^*(\xi)\Big)(1+\epsilon^+ v) \quad \forall \; v \in [0,\hat{\delta}]
\end{eqnarray}
\begin{eqnarray}
    \underline{\beta}(v) \equiv \Big(\min_{\xi \in [0,v]} \beta^*(\xi)\Big)(1-\epsilon^- v) \quad \forall \; v \in [0,\hat{\delta}]
\end{eqnarray}
where $\epsilon^+>0$ and $\epsilon^->0$ are chosen to be small enough such that
\begin{eqnarray}
    1 < \underline{\beta}(\hat{\delta}) < \overline{\beta}(\hat{\delta}) < \inf_{v \in (0,\hat{\delta}]} \frac{\Phi^{-1}(v)}{v}.
\end{eqnarray}
Notice then that $\overline{\beta},\underline{\beta}$ are both continuous. Also, $\overline{\beta}$ is stirctly increasing and $\underline{\beta}$ is strictly decreasing on $[0,\hat{\delta}]$. Finally, notice that
\begin{eqnarray}
    \underline{\beta}(0) = \beta^*(0) = \frac{2N}{N+1} = \overline{\beta}(0)
\end{eqnarray}
\begin{eqnarray}
    \underline{\beta}(v) < \beta^*(v), \frac{2N}{N+1} < \overline{\beta}(v) \quad \forall \; v \in (0,\hat{\delta}].
\end{eqnarray}

We then define the space $\mathcal{M}([0,\hat{\delta}])$ of all continuous functions $\gamma:[0,\hat{\delta}] \rightarrow \mathbb{R}$ such that $\gamma(v) \in [\underline{\beta}(v),\overline{\beta}(v)]$ for all $v \in [0,\hat{\delta}]$.

It is also useful to define the larger space $\mathcal{N}([0,\hat{\delta}])$ of all continuous functions $\gamma:[0,\hat{\delta}] \rightarrow \mathbb{R}$ such that $\gamma(0) = 2N/(N+1)$. It is clear that $\mathcal{M}([0,\hat{\delta}]) \subseteq \mathcal{N}([0,\hat{\delta}])$.

{\bf Step \ref{step: local existence}-c}: Consider the set $\mathcal{S} \equiv \{(\beta,v): v\in(0,\hat{\delta}],\beta \in [\underline{\beta}(v),\overline{\beta}(v)]\}$. Then, there exists a constant $\bar{K}_{\beta}$ such that $|K_\beta(\beta,v)|\leq \bar{K}_\beta < \infty$ for all $(\beta,v) \in \mathcal{S}$.

To prove this, we will establish the boundedness of each component of (\ref{equation: K beta partial derivative}). Firstly, $f(v)v/F(v)$ is continuous on $(0,\bar{v}]$ with $\lim_{v \downarrow 0} f(v)v/F(v) = 1$, so it is clearly bounded on $\mathcal{S}$.

Secondly, notice that $\beta - 1 \geq \underline{\beta}(\hat{\delta})-1 > 0$ for all $(\beta,v) \in \mathcal{S}$. Also, from the assumptions on $g$, we can find constants $\underline{g},\overline{g}>0$ such that $g(w) \in [\underline{g},\overline{g}]$ for all $w \in [0,\bar{w}]$. Hence it is clear that we may bound $(\beta-1)g(\beta v)$ away from 0 on $(\beta,v) \in \mathcal{S}$.

Thirdly, $G(\beta v) g(\beta v)/v \leq \overline{g} \beta v \frac{\overline{g}}{v} \leq \overline{g}^2 \overline{\beta}(\hat{\delta})$ for all $(\beta,v) \in \mathcal{S}$.

Fourthly, notice that
\begin{eqnarray}
    \Big(\frac{1}{v}\int_0^{\beta v} \xi g(\xi) \,d\xi\Big)\frac{g(\beta v)}{v} \leq \Big(\frac{1}{v}\overline{g} \int_0^{\beta v} \xi \,d\xi\Big) \frac{\overline{g}}{v} \leq \Big(\frac{\overline{g}}{v}\Big)^2\frac{1}{2}(\beta v)^2 \leq \frac{1}{2} \overline{g}^2 \overline{\beta}(\hat{\delta})^2
\end{eqnarray}
for all $(\beta,v) \in \mathcal{S}$.

Fifthly, from the continuity of $g'$, it follows that $g'$ is bounded on $[0,\bar{w}]$.

Combining the above, we can clearly see the boundedness of $K_{\beta}(\beta,v)$ on $\mathcal{S}.$

{\bf Step \ref{step: local existence}-d}: We now define mappings for which we will establish our fixed point arguments.

We first define $\boldsymbol{\mathcal{T}}: \mathcal{M}([0,\hat{\delta}]) \rightarrow \mathcal{N}([0,\hat{\delta}])$ as below. Given $\gamma \in \mathcal{M}([0,\hat{\delta}])$, define $\boldsymbol{\mathcal{T}}\gamma:[0,\hat{\delta}] \rightarrow \mathbb{R}$ as below:
\begin{eqnarray}
    \boldsymbol{\mathcal{T}} \gamma(v) = \frac{1}{v} \int_0^v K(\gamma(\xi),\xi)\,d\xi \quad \forall \; v \in (0,\hat{\delta}]
\end{eqnarray}
\begin{eqnarray}
    \boldsymbol{\mathcal{T}}\gamma(0) = \frac{2N}{N+1}.
\end{eqnarray}
It remains to be established that $\boldsymbol{\mathcal{T}}\gamma$ defined as above belongs to $\mathcal{N}([0,\hat{\delta}])$. In particular, we must establish the continuity of $\boldsymbol{\mathcal{T}} \gamma$ at 0. This is done below. Firstly, by an application of L'Hopital's rule,
\begin{eqnarray}
    \lim_{v \downarrow 0} \frac{1}{v} \int_0^v K(\gamma(\xi),\xi)\,d\xi
 = \lim_{v \downarrow 0} K(\gamma(v),v).
 \end{eqnarray}
Then, we have that the above is equal to
\begin{eqnarray}
    \lim_{v \downarrow 0} [K(\gamma(0),v) + \{K(\gamma(v),v) - K(\gamma(0),v)\}].
\end{eqnarray}
Since $\gamma \in \mathcal{M}([0,\hat{\delta}])$, we have $\gamma(0) = 2N/(N+1)$. Then, by Step 3-b, we know that $\lim_{v \downarrow 0} K(\gamma(0),v) = 2N/(N+1)$. Furthermore, by Step 3-c, we know that
\begin{eqnarray}
    |K(\gamma(v),v) - K(\gamma(0),v)| \leq \bar{K}_\beta |\gamma(v)-\gamma(0)|,
\end{eqnarray}
which converges to 0 as $v \downarrow 0$ due to the continuity of $\gamma$. Therefore, it follows that
\begin{eqnarray}
    \lim_{v \downarrow 0} \boldsymbol{\mathcal{T}}\gamma(v) = \boldsymbol{\mathcal{T}}\gamma(0) = \frac{2N}{N+1},
\end{eqnarray}
which establishes our desired claim of continuity at 0. Thus, we have a well-defined mapping $\boldsymbol{\mathcal{T}}:\mathcal{M}([0,\hat{\delta}]) \rightarrow \mathcal{N}([0,\hat{\delta}])$.

We then define the mapping $\boldsymbol{\mathcal{U}}: \mathcal{M}([0,\hat{\delta}]) \rightarrow \mathcal{M}([0,\hat{\delta}])$ as below. Given $\gamma \in \mathcal{M}([0,\hat{\delta}])$, define the function $\boldsymbol{\mathcal{U}}\gamma:[0,\hat{\delta}] \rightarrow \mathbb{R}$ as below:
\begin{eqnarray}
    \boldsymbol{\mathcal{U}} \gamma(v) =
    \begin{cases} \underline{\beta}(v) \quad {\rm if} \; \boldsymbol{\mathcal{T}}\gamma(v) < \underline{\beta}(v) \\
    \boldsymbol{\mathcal{T}} \gamma(v) \quad {\rm if } \; \underline{\beta}(v) \leq \boldsymbol{\mathcal{T}}\gamma(v) \leq \overline{\beta}(v) \\
    \overline{\beta}(v) \quad {\rm if } \; \boldsymbol{\mathcal{T}}\gamma(v) > \overline{\beta}(v)
    \end{cases}
\end{eqnarray}
By construction it is clear that $\boldsymbol{\mathcal{U}}\gamma \in \mathcal{M}([0,\hat{\delta}])$.

{\bf Step \ref{step: local existence}-e}: Under the sup-norm, $\boldsymbol{\mathcal{U}}$ is a continuous mapping.

Let $\gamma_1,\gamma_2 \in \mathcal{M}([0,\hat{\delta}])$. Then,
\begin{eqnarray}
\begin{split}
    ||\boldsymbol{\mathcal{U}}\gamma_1 - \boldsymbol{\mathcal{U}} \gamma_2|| & = \max_{v \in [0,\hat{\delta}]} |\boldsymbol{\mathcal{U}}\gamma_1(v) - \boldsymbol{\mathcal{U}}\gamma_2(v)| \leq \max_{v \in [0,\hat{\delta}]} |\boldsymbol{\mathcal{T}}\gamma_1(v) - \boldsymbol{\mathcal{T}}\gamma_2(v)| \\
     & = \sup_{v \in (0,\hat{\delta}]} \Big| \frac{1}{v} \int_0^v [K(\gamma_1(\xi),\xi)-K(\gamma_2(\xi),\xi)] \,d\xi \Big| \\
     & \leq \sup_{v\in (0,\hat{\delta}]} \frac{1}{v} \int_0^v |K(\gamma_1(\xi),\xi)-K(\gamma_2(\xi),\xi)|\,d\xi.
\end{split}
\end{eqnarray}
We may then use step 3-c to obtain
\begin{eqnarray}
\begin{split}
    ||\boldsymbol{\mathcal{U}}\gamma_1 - \boldsymbol{\mathcal{U}} \gamma_2|| & \leq \sup_{v \in (0,\hat{\delta}]} \frac{1}{v} \int_0^v \bar{K}_\beta |\gamma_1(\xi)-\gamma_2(\xi)|\,d\xi \\
     & \leq \bar{K}_\beta || \gamma_1-\gamma_2||,
\end{split}
\end{eqnarray}
from which continuity of $\boldsymbol{\mathcal{U}}$ follows.

{\bf Step \ref{step: local existence}-f}: The image $\boldsymbol{\mathcal{U}}(\mathcal{M}([0,\hat{\delta}]))$ of $\boldsymbol{\mathcal{U}}$ is a uniformly equicontinuous family of functions.

First, we show that there is a constant $\bar{K}<\infty$ such that for all $(\beta,v)\in \mathcal{S}$, we have
\begin{eqnarray}\label{equation: bound on K}
    |K(\beta,v)| \leq \bar{K}.
\end{eqnarray}
To see this, first note that $\lim_{v \downarrow 0} K(2N/(N+1),v)$ is a finite quantity and hence that $\{K(2N/(N+1),v): v \in (0,\hat{\delta}]\}$ is a bounded set. Then, using Step 3-c, we can show that
\begin{eqnarray}
\begin{split}
    |K(\beta,v)| & \leq \Big|K\Big(\frac{2N}{N+1},v\Big)\Big| + \Big|K(\beta,v)-K\Big(\frac{2N}{N+1},v\Big)\Big| \\
    & \leq \Big|K\Big(\frac{2N}{N+1},v\Big)\Big| + \bar{K}_\beta|\beta - \frac{2N}{N+1}| \\
    & \leq \Big|K\Big(\frac{2N}{N+1},v\Big)\Big| + \bar{K}_\beta|\overline{\beta}(\hat{\delta})-\underline{\beta}(\hat{\delta})|
\end{split}
\end{eqnarray}
for all $(\beta,v) \in \mathcal{S}$, which clearly establishes our claim.

Now we proceed to establish uniform equicontinuity. Let us fix an arbitrary $\eta>0$.

Firstly, since $\lim_{v \downarrow 0} [\overline{\beta}(v)-\underline{\beta}(v)]=0$, we can find some $\xi_1 \in (0,\hat{\delta}]$ such that $\overline{\beta}(\xi_1)-\underline{\beta}(\xi_1) < \eta/2$.

Secondly, by the Heine-Cantor theorem, both $\overline{\beta}$ and $\underline{\beta}$ are uniformly continuous on $[\xi_1,\hat{\delta}]$. Thus, we can find some $\xi_2>0$ such that for every $v_1,v_2 \in [\xi_1,\hat{\delta}]$ with $|v_1-v_2|<\xi_2$,
\begin{eqnarray}\label{equation: uniform continuity upper beta}
    |\overline{\beta}(v_1)-\overline{\beta}(v_2)| < \frac{\eta}{2}
\end{eqnarray}
\begin{eqnarray}\label{equation: uniform continuity lower beta}
    |\underline{\beta}(v_1)-\underline{\beta}(v_2)| < \frac{\eta}{2}.
\end{eqnarray}

Thirdly, given $v_1,v_2 \in [\xi_1,\hat{\delta}]$, notice that for any $\gamma \in \mathcal{M}([0,\hat{\delta}])$
\begin{eqnarray}
\begin{array}{c}
    |\boldsymbol{\mathcal{T}}\gamma(v_1)-\boldsymbol{\mathcal{T}}\gamma(v_2)| = \bigg|\frac{\int_0^{v_1} K(\gamma(v),v)\,dv}{v_1} - \frac{\int_0^{v_2} K(\gamma(v),v)\,dv}{v_2} \bigg| \\
    = \bigg|\frac{\int_0^{v_1} K(\gamma(v),v)\,dv}{v_1} - \frac{\int_0^{v_2} K(\gamma(v),v)\,dv}{v_1} + \frac{\int_0^{v_2} K(\gamma(v),v)\,dv}{v_1} - \frac{\int_0^{v_2} K(\gamma(v),v)\,dv}{v_2} \bigg|.
\end{array}
\end{eqnarray}
Then, using (\ref{equation: bound on K}), we can establish that
\begin{eqnarray}
\begin{split}
    |\boldsymbol{\mathcal{T}}\gamma(v_1)-\boldsymbol{\mathcal{T}}\gamma(v_2)| & \leq \frac{1}{\xi_1} \bar{K} |v_1-v_2| + \bar{K} v_2 \Big|\frac{v_1-v_2}{v_1 v_2}\Big| \\
    & \leq \frac{2}{\xi_2} \bar{K} |v_1-v_2|.
\end{split}
\end{eqnarray}
From this it is clear that we can set $\xi_3>0$ such that for every $v_1,v_2 \in [\xi_1,\hat{\delta}]$ with $|v_1-v_2|<\xi_3$, we have for every $\gamma \in \mathcal{M}([0,\hat{\delta}])$ that
\begin{eqnarray}\label{inequality: T difference bound}
    |\boldsymbol{\mathcal{T}}\gamma(v_1)-\boldsymbol{\mathcal{T}}\gamma(v_2)| < \frac{\eta}{2}.
\end{eqnarray}

Let us now define $\xi^* \equiv \min \{\xi_2,\xi_3\}$. Then, fix any $\gamma \in \mathcal{M}([0,\hat{\delta}])$ and $v_1,v_2 \in [0,\hat{\delta}]$ with $|v_1-v_2|<\xi^*$. To establish uniform equicontinuity, it suffices to show that $|\boldsymbol{\mathcal{U}} \gamma(v_1) - \boldsymbol{\mathcal{U}} \gamma(v_2)|<\eta$. We do this by considering three separate cases.

\underline{Case 1: $v_1,v_2 \in [0,\xi_1]$}.

Then, it is immediately clear from the construction of $\overline{\beta},\underline{\beta}$ that
\begin{eqnarray}
    |\boldsymbol{\mathcal{U}} \gamma(v_1) - \boldsymbol{\mathcal{U}} \gamma(v_2)|\leq \overline{\beta}(\xi_1) - \underline{\beta}(\xi_1) < \frac{\eta}{2}.
\end{eqnarray}

\underline{Case 2: $v_1,v_2 \in [\xi_1,\hat{\delta}]$}.

Without loss of generality, let $v_1<v_2$. Then, we can analyze the four sub-cases below:

\textit{Case 2(a): $\boldsymbol{\mathcal{T}}\gamma(v_1) \in [\underline{\beta}(v_1),\overline{\beta}(v_1)]$.}

By construction of the functions $\overline{\beta},\underline{\beta}$ it is then clear that 
\begin{eqnarray}\label{inequality: U-T}
    |\boldsymbol{\mathcal{U}} \gamma(v_1) - \boldsymbol{\mathcal{U}} \gamma(v_2)| \leq |\boldsymbol{\mathcal{T}}\gamma(v_1) - \boldsymbol{\mathcal{T}}\gamma(v_2)| < \frac{\eta}{2}.
\end{eqnarray}

\textit{Case 2(b): Either (1) $\boldsymbol{\mathcal{T}}\gamma(v_1)<\underline{\beta}(v_1)$ and $\boldsymbol{\mathcal{T}}\gamma(v_2)< \underline{\beta}(v_2)$; or (2) $\boldsymbol{\mathcal{T}}\gamma(v_1)>\overline{\beta}(v_1)$ and $\boldsymbol{\mathcal{T}}\gamma(v_2)>\overline{\beta}(v_2)$.}

Under the first case, it is clear that
\begin{eqnarray}
    |\boldsymbol{\mathcal{U}}\gamma(v_1)-\boldsymbol{\mathcal{U}}\gamma(v_2)| = |\underline{\beta}(v_1)-\underline{\beta}(v_2)| < \frac{\eta}{2}.
\end{eqnarray}
An analogous inequality holds for the second case as well.

\textit{Case 2(c): Either (1) $\boldsymbol{\mathcal{T}}\gamma(v_1)<\underline{\beta}(v_1)$ and $\boldsymbol{\mathcal{T}}\gamma(v_2)> \overline{\beta}(v_2)$; or (2) $\boldsymbol{\mathcal{T}}\gamma(v_1)>\overline{\beta}(v_1)$ and $\boldsymbol{\mathcal{T}}\gamma(v_2)<\underline{\beta}(v_2)$.}

Under both cases, it is clear that by construction the inequality (\ref{inequality: U-T}) must hold.

\textit{Case 2(d): Either (1) $\boldsymbol{\mathcal{T}}\gamma(v_1)<\underline{\beta}(v_1)$ and $\boldsymbol{\mathcal{T}}\gamma(v_2) \in [\underline{\beta}(v_2),\overline{\beta}(v_2)]$; or (2) $\boldsymbol{\mathcal{T}}\gamma(v_1)>\overline{\beta}(v_1)$ and $\boldsymbol{\mathcal{T}}\gamma(v_2) \in [\underline{\beta}(v_2),\overline{\beta}(v_2)]$.}

Consider the first case. Due to (\ref{inequality: T difference bound}), we have that 
\begin{eqnarray}
    \boldsymbol{\mathcal{T}}\gamma(v_2) < \boldsymbol{\mathcal{T}}\gamma(v_1) + \frac{\eta}{2} < \underline{\beta}(v_1) + \frac{\eta}{2}.
\end{eqnarray}
Then, notice that 
\begin{eqnarray}
    \boldsymbol{\mathcal{U}}\gamma(v_2) - \boldsymbol{\mathcal{U}} \gamma(v_1) = \boldsymbol{\mathcal{T}}\gamma(v_2) - \underline{\beta}(v_1) < \frac{\eta}{2}.
\end{eqnarray}
Also, we have that
\begin{eqnarray}
    \boldsymbol{\mathcal{U}}\gamma(v_2) - \boldsymbol{\mathcal{U}} \gamma(v_1) \geq \underline{\beta}(v_2) - \underline{\beta}(v_1) > -\frac{\eta}{2}.
\end{eqnarray}
Therefore, it is clear that $|\boldsymbol{\mathcal{U}}\gamma(v_2) - \boldsymbol{\mathcal{U}}\gamma(v_1)| < \eta/2$. An analogous argument can be established under the second case as well.

\underline{Case 3: $v_1 \in [0, \xi_1]$ and $v_2 \in [\xi_1,\hat{\delta}]$}.

We then have by combining Cases 1 and 2 above that
\begin{eqnarray}
    |\boldsymbol{\mathcal{U}}\gamma(v_1) - \boldsymbol{\mathcal{U}}\gamma(v_2)| \leq |\boldsymbol{\mathcal{U}}\gamma(v_1) - \boldsymbol{\mathcal{U}}\gamma(\xi_1)| + |\boldsymbol{\mathcal{U}} \gamma(\xi_1) - \boldsymbol{\mathcal{U}} \gamma(v_2)| < \frac{\eta}{2} + \frac{\eta}{2} = \eta.
\end{eqnarray}

{\bf Step \ref{step: local existence}-g}: The mapping $\boldsymbol{\mathcal{U}}$ has a fixed point $\gamma^*$.

Consider the space of continuous functions mapping $[0,\hat{\delta}]$ to $\mathbb{R}$, which is a normed linear space under the sup-norm. Then, notice $\mathcal{M}([0,\hat{\delta}])$ is a nonempty closed convex subset of this normed linear space.

Furthermore, we know that the image $\boldsymbol{\mathcal{U}}(\mathcal{M}([0,\hat{\delta}]))$ of $\boldsymbol{\mathcal{U}}$ is uniformly equicontinuous by step 3-f. Also, it is clear by construction that $\boldsymbol{\mathcal{U}}(\mathcal{M}([0,\hat{\delta}]))$ is uniformly bounded. Hence it follows from the Arzel\`a-Ascoli theorem that $\boldsymbol{\mathcal{U}}(\mathcal{M}([0,\hat{\delta}]))$ is relatively compact. 

Finally, recall that $\boldsymbol{\mathcal{U}}$ is continuous by step 3-e. Therefore, by Schauder's theorem, $\boldsymbol{\mathcal{U}}$ must have a fixed point $\gamma^*$.

{\bf Step \ref{step: local existence}-h}: $\gamma^*$ is also a fixed point of $\boldsymbol{\mathcal{T}}$.

It suffices to prove that $\underline{\beta}(v)\leq \boldsymbol{\mathcal{T}}\gamma^*(v) \leq \overline{\beta}(v)$ for all $v \in [0,\hat{\delta}]$. We will prove below that $\boldsymbol{\mathcal{T}}\gamma^*(v) \leq \overline{\beta}(v)$ for all $v \in [0,\hat{\delta}]$. The proof for the other inequality follows a symmetric argument.

Suppose for contradiction that $\boldsymbol{\mathcal{T}}\gamma^*(\hat{v})>\overline{\beta}(\hat{v})$ for some 
$\hat{v} \in (0,\hat{\delta}]$. We can then show that there exists some $\tilde{v} \in (0, \hat{v})$ such that $\boldsymbol{\mathcal{T}}\gamma^*(\tilde{v})<\overline{\beta}(\tilde{v})$. To see this, suppose that $\boldsymbol{\mathcal{T}}\gamma^*(v) \geq \overline{\beta}(v)$ for all $v \in (0,\hat{v})$. Then, we would have that
\begin{eqnarray}
    \gamma^*(v) = \boldsymbol{\mathcal{U}} \gamma^*(v) = \overline{\beta}(v) \quad \forall \; v \in (0,\hat{v}],
\end{eqnarray}
which in turn implies
\begin{eqnarray}
    K(\gamma^*(v),v) < \frac{2N}{N+1} \quad \forall \; v \in (0,\hat{v}]
\end{eqnarray}
and hence
\begin{eqnarray}
    \boldsymbol{\mathcal{T}}\gamma^*(\hat{v}) = \frac{1}{\hat{v}}\int_0^{\hat{v}} K(\gamma^*(v),v) \,dv \leq \frac{2N}{N+1} < \overline{\beta}(\hat{v}),
\end{eqnarray}
which is clearly a contradiction.

Let us then define $E \equiv \{v \in [\tilde{v},\hat{v}]: \boldsymbol{\mathcal{T}} \gamma^*(v) > \overline{\beta}(v)\}$, which must clearly be non-empty. Hence define $v^* \equiv \inf E$, which clearly lies in the interval $(\tilde{v},\hat{v})$. Furthermore, by continuity, it must be that $\boldsymbol{\mathcal{T}}\gamma^*(v^*) = \overline{\beta}(v^*)$. However, notice that
\begin{eqnarray}
    K(\gamma^*(v^*),v^*) = K(\overline{\beta}(v^*),v^*)<\frac{2N}{N+1}
\end{eqnarray}
while $\gamma^*(v^*) = \overline{\beta}(v^*) > 2N/(N+1)$. It is then clear that $\gamma^*$ is strictly decreasing in some right-neighborhood of $v^*$. However, since $\overline{\beta}$ is strictly increasing, we reach a contradiction of $v^*$ as the infimum of $E$. Consequently, we establish step 3-h.

{\bf Step \ref{step: local existence}-i}: We now complete the proof of Step 3.

Let us define
\begin{eqnarray}
    b(v) = \gamma^*(v)v \quad \forall \; v \in [0,\hat{\delta}]
\end{eqnarray}
Then, we have by step 3-h that for all $v \in (0,\hat{\delta}]$,
\begin{eqnarray}
    b(v) = \boldsymbol{\mathcal{T}}\gamma^*(v) v = \int_0^v K(\gamma^*(\xi),\xi)\,d\xi = \int_0^v H(b(\xi),\xi)\,d\xi.
\end{eqnarray}
The claim of Step \ref{step: local existence} is then clear.

\begin{step}\label{step: global existence}
The local solution $b:[0,\hat{\delta}]\rightarrow \mathbb{R}_+$ constructed in Step 3 can be extended to a global solution $\hat{b}: [0,\bar{v}] \rightarrow \mathbb{R}_+$ of the ODE (\ref{equation: ODE}).
\end{step}

It suffices to show the existence of a function $\bar{b}: [\hat{\delta},\bar{v}] \rightarrow \mathbb{R}_+$ such that (1) $\bar{b}(\hat{\delta}) = b(\hat{\delta})$, and (2) $\bar{b}$ satisfies the ODE (\ref{equation: ODE}) on $[\hat{\delta},\bar{v}]$. Notice that this corresponds to a standard Cauchy problem.

Consider the space $A \equiv \{(v,b): v\in [\hat{\delta},\bar{v}),b \in (v,\Phi^{-1}(v))\}$, and notice that the function $H$ from (\ref{equation: ODE}) is well-defined on this restricted space $A$. Furthermore, it is clear that $H$ is continuous on $A$. Also, notice that $\partial H(b,v)/\partial b$ exists and is continuous on $A$. Consequently, it follows from the Cauchy-Lipschitz theorem that (1) given any point $(\hat{\delta},b) \in A$, there is a unique local right solution for the ODE (\ref{equation: ODE}), and (2) given any point $(v,b)\in A$ with $v \in (\hat{\delta},\bar{v})$, there is a unique local solution for the ODE (\ref{equation: ODE}).

Let us then define the set $\phi$ as the collection of all $v \in (\hat{\delta},\bar{v})$ for which there exists $\tilde{b}:[\hat{\delta},v) \rightarrow \mathbb{R}_+$ such that $\tilde{b}(\hat{\delta})=b(\hat{\delta})$ and $\tilde{b}$ satisfies (\ref{equation: ODE}) on its domain $[\hat{\delta},v)$. Then, we know that $\phi$ is nonempty with $\sup \phi \in (\hat{\delta},\bar{v}]$ due to the local existence property established above. Furthermore, we know that if $v \in [\hat{\delta},\sup \phi)$, every possible function $\tilde{b}:[\hat{\delta},v^+)\rightarrow \mathbb{R}_+$ with $v^+ \in (v,\sup \phi]$ and $\tilde{b}(\hat{\delta}) = b(\hat{\delta})$ and $\tilde{b}$ satisfying (\ref{equation: ODE}) on $[\hat{\delta},v^+)$ must agree on the value of $\tilde{b}(v)$. Thus, we may construct a uniquely well-defined function $\tilde{b}:[\hat{\delta},\sup \phi) \rightarrow \mathbb{R}_+$. It is immediately clear that $\tilde{b}(\hat{\delta}) = b(\hat{\delta})$ and $\tilde{b}$ satisfies (\ref{equation: ODE}) on $[\hat{\delta},\sup \phi)$.

Since $H(b,v)>0$ for all $(b,v) \in A$, we must have that $\tilde{b}$ defined above is strictly increasing on $[\hat{\delta},\sup \phi)$. Hence $\lim_{v \uparrow \sup \phi} \tilde{b}(v)$ exists. We will now prove that
\begin{eqnarray}\label{equation: limit interval}
    \lim_{v \uparrow \sup \phi} \tilde{b}(v) \in (\sup \phi, \Phi^{-1}(\sup \phi))
\end{eqnarray}
by ruling out the two cases shown below:

\vspace{2mm}

\underline{Case 1: Suppose $\lim_{v \uparrow \sup \phi} \tilde{b}(v) = \sup \phi$.}

Then, we clearly have
\begin{eqnarray}
    \lim_{v \uparrow \sup \phi} H(\tilde{b}(v),v) = + \infty,
\end{eqnarray}
which in turn implies that there is some $\tilde{v}< \sup \phi$ such that $\tilde{b}(\tilde{v})<\tilde{v}$. This constitutes a contradiction.

\underline{Case 2: Suppose $\lim_{v \uparrow \sup \phi} \tilde{b}(v) = \Phi^{-1}(\sup \phi)$.}

Then, we have
\begin{eqnarray}
    \lim_{v \uparrow \sup \phi} H(\tilde{b}(v),v) = 0 < \frac{d \Phi^{-1}(\sup \phi)}{dv}
\end{eqnarray}
where the inequality follows from step \ref{step: Phi}. However, it is then clear that there is some $\tilde{v}<\sup \phi$ such that $\tilde{b}(\tilde{v})>\Phi^{-1}(\tilde{v}),$ which is once again a contradiction.

\vspace{2mm}

We will now proceed to prove that $\sup \phi = \bar{v}$. Suppose for contradiction that $\sup \phi < \bar{v}$. Then, due to (\ref{equation: limit interval}), we can extend the solution $\tilde{b}$ to a neighborhood of $\sup \phi$. However, this contradicts the definition of $\sup \phi$.

We therefore have a function $\tilde{b}:[\hat{\delta},\bar{v}) \rightarrow \mathbb{R}_+$ with $\tilde{b}(\hat{\delta}) = b(\hat{\delta})$ and $\tilde{b}$ satisfying (\ref{equation: ODE}) on $[\hat{\delta},\bar{v})$. We also know that $\lim_{v \uparrow \bar{v}} \tilde{b}(v) \in (\bar{v},\Phi^{-1}(\bar{v}))$. Let us then define
\begin{eqnarray}
    \bar{b}(v) =
    \begin{cases}
        \tilde{b}(v) \quad {\rm if} \; v \in [\hat{\delta},\bar{v}) \\
        \lim_{v \uparrow \bar{v}} \tilde{b}(v) \quad {\rm if} \; v = \bar{v}
    \end{cases}
\end{eqnarray}
Then, to complete the proof of step \ref{step: global existence} it suffices to show that
\begin{eqnarray}\label{equation: ODE at v-bar}
    \frac{\partial \bar{b}(\bar{v})}{\partial v} = H(\bar{b}(\bar{v}),\bar{v}).
\end{eqnarray}
Notice that for any $v \in [\hat{\delta},\bar{v})$, we must have
\begin{eqnarray}
    \frac{\bar{b}(\bar{v})-\bar{b}(v)}{\bar{v}-v} = \bar{b}'(v^*) = H(\tilde{b}(v^*),v^*)
\end{eqnarray}
for some $v^* \in (v,\bar{v})$ due to the Mean Value Theorem. Then, since
\begin{eqnarray}
    \lim_{v \uparrow \bar{v}} H(\tilde{b}(v),v) = H(\bar{b}(\bar{v}),\bar{v}),
\end{eqnarray}
it is clear that (\ref{equation: ODE at v-bar}) must hold.

\begin{step}
We conclude the proof of existence by establishing that the $\hat{b}$ constructed in step \ref{step: global existence} induces an equilibrium in the tournament auction.
\end{step}

Consider a first-stage bidder with value $v \in [0,\bar{v}]$. Given that (1) all other first-stage bidders bid according to $\hat{b}$, and (2) the second-stage bidder bids his value, we must show that the optimal bid is $\hat{b}(v)$.

Firstly, if $v=0$, then the optimality of $\hat{b}(0) = 0$ is obvious.

Let $v \in (0,\bar{v}]$. We first show that it is suboptimal to bid $b > \hat{b}(\bar{v})$. In particular, it is strictly better to submit the bid $\hat{b}(\bar{v})$. To see this, note that the payoff under bid $b$ is $\int_0^b (v-\xi)g(\xi)\,d\xi$, whereas the payoff under bid $\hat{b}(\bar{v})$ is $\int_0^{\hat{b}(\bar{v})} (v-\xi)g(\xi)\,d\xi$. Since $\hat{b}(\bar{v})>\bar{v} \geq v$, it is immediately clear that bidding $\hat{b}(\bar{v})$ is strictly better than bidding $b$.

Hence we have reduced the bidder's problem to choosing a bid from $[0,\hat{b}(\bar{v})]$. This is equivalent to the set $\{\hat{b}(\tilde{v}): \tilde{v} \in [0,\bar{v}]\}$. Hence we can reformulate the bidder's problem as choosing a \textit{reported value} $\tilde{v}$ as below:
\begin{eqnarray}
    \max_{\tilde{v} \in [0,\bar{v}]} \bigg\{\pi(\tilde{v}|v) \equiv F(\tilde{v})^{N-1} \int_0^{\hat{b}(\tilde{v})} (v-\xi)g(\xi) \,d\xi \bigg\}
\end{eqnarray}

We may then compute the derivative
\begin{eqnarray}
    \frac{\partial \pi(\tilde{v}|v)}{\partial \tilde{v}} = (N-1)F(\tilde{v})^{N-2}f(\tilde{v}) \int_0^{\hat{b}(\tilde{v})} (v-\xi) g(\xi) \,d\xi + F(\tilde{v})^{N-1}(v-\hat{b}(\tilde{v}))g(\hat{b}(\tilde{v})) \frac{\partial \hat{b}(\tilde{v})}{\partial \tilde{v}}.
\end{eqnarray}
Notice first that
\begin{eqnarray}\label{equation: sign below self-cutoff}
    \frac{\partial \pi(\tilde{v}|v)}{\partial \tilde{v}}>0 \quad {\rm for} \; \tilde{v} \in (0,\hat{b}^{-1}(v)]. 
\end{eqnarray}
Then, for $\tilde{v} \in (\hat{b}^{-1}(v),\bar{v}]$, notice that we have
\begin{eqnarray}
\begin{split}
    \frac{\partial \pi(\tilde{v}|v)}{\partial \tilde{v}} = & F(\tilde{v})^{N-1} (\hat{b}(\tilde{v})-v) g(\hat{b}(\tilde{v})) \cdot \\
    & \cdot \Big[(N-1)\frac{f(\tilde{v})}{F(\tilde{v})} \frac{1}{\hat{b}(\tilde{v})-v} \frac{G(\hat{b}(\tilde{v}))}{g(\hat{b}(\tilde{v}))} \big(v-\Phi(\hat{b}(\tilde{v}))\big) - \frac{\partial \hat{b}(\tilde{v})}{\partial \tilde{v}} \Big],
\end{split}
\end{eqnarray}
from which it follows that
\begin{eqnarray}
    {\rm sign} \Big(\frac{\partial \pi(\tilde{v}|v)}{\partial \tilde{v}}\Big) = {\rm sign} \Big((N-1)\frac{f(\tilde{v})}{F(\tilde{v})} \frac{1}{\hat{b}(\tilde{v})-v} \frac{G(\hat{b}(\tilde{v}))}{g(\hat{b}(\tilde{v}))} \big(v-\Phi(\hat{b}(\tilde{v}))\big) - \frac{\partial \hat{b}(\tilde{v})}{\partial \tilde{v}} \Big).
\end{eqnarray}
Then, we can substitute (\ref{equation: ODE}) and simplify to obtain
\begin{eqnarray}
    {\rm sign} \Big(\frac{\partial \pi(\tilde{v}|v)}{\partial \tilde{v}}\Big) = {\rm sign} \Big( \frac{v-\Phi(\hat{b}(\tilde{v}))}{\hat{b}(\tilde{v})-v} - \frac{\tilde{v}-\Phi(\hat{b}(\tilde{v}))}{\hat{b}(\tilde{v})-\tilde{v}} \Big).
\end{eqnarray}
We can then see that
\begin{eqnarray}\label{equation: sign above self-cutoff}
    \frac{\partial \pi(\tilde{v}|v)}{\partial \tilde{v}}
    \begin{cases}
        >0 \quad {\rm if} \; \tilde{v} \in (\hat{b}^{-1}(v),v)\\
        =0 \quad {\rm if} \; \tilde{v} = v \\
        <0  \quad {\rm if} \; \tilde{v} \in (v,\bar{v}] 
    \end{cases}
\end{eqnarray}
Thus, from (\ref{equation: sign below self-cutoff}) and (\ref{equation: sign above self-cutoff}), it is clear that the bid $\hat{b}(v)$ is optimal. 
This then establishes that $\hat{b}$ induces an equilibrium in the tournament auction as required.

\begin{step}
We now establish that any $b(\cdot)$ inducing an equilibrium must be strictly increasing.
\end{step}

We begin by first establishing that any such $b(\cdot)$ must be weakly increasing. Suppose by way of contradiction that we can find $v_1,v_2 \in [0,\bar{v}]$ such that $v_1 < v_2$ and $b(v_1) > b(v_2)$. Fix any first-stage bidder $i$. Let us denote as $P_i(b_i;b(\cdot))$ the probability of $i$ winning the first stage given bid $b_i$ and that all others bid according to the equilibrium bidding function $b(\cdot)$.\footnote{Notice that we allow $P_i(b_i;b(\cdot))$ to be bidder-specific due to potentially unequal tie-breaking rules. However, once we prove this proposition, it will be clear that ties arise with zero probability in equilibrium and tie-breaking rules do not matter.} Then, from optimality of the equilibrium bids, we must have
\begin{eqnarray}
    P_i(b(v_1);b(\cdot)) \int_0^{b(v_1)} (v_1 - \xi) g(\xi) \,d\xi \geq P_i(b(v_2);b(\cdot)) \int_0^{b(v_2)} (v_1 - \xi) g(\xi) \,d\xi.
\end{eqnarray}
Then, it is clear that
\begin{eqnarray}
\begin{split}
    & P_i(b(v_1);b(\cdot)) \int_0^{b(v_1)} (v_2 - \xi) g(\xi) \,d\xi \\ & \geq P_i(b(v_2);b(\cdot)) \int_0^{b(v_2)} (v_1 - \xi) g(\xi) \,d\xi + P_i(b(v_1);b(\cdot)) \int_0^{b(v_1)} (v_2-v_1) g(\xi) \, d\xi.
\end{split}
\end{eqnarray}
Then, since $b(v_1)>b(v_2)$ by assumption, we have
\begin{eqnarray}
\begin{split}
    & P_i(b(v_1);b(\cdot)) \int_0^{b(v_1)} (v_2 - \xi) g(\xi) \,d\xi \\ & > P_i(b(v_2);b(\cdot)) \int_0^{b(v_2)} (v_1 - \xi) g(\xi) \,d\xi + P_i(b(v_2);b(\cdot)) \int_0^{b(v_2)} (v_2-v_1) g(\xi) \, d\xi \\
    & = P_i(b(v_2);b(\cdot)) \int_0^{b(v_2)} (v_2 - \xi) g(\xi) \,d\xi.
\end{split}
\end{eqnarray}
However, this contradicts the optimality of bid $b(v_2)$ given value $v_2$.

The argument above has established that $b(\cdot)$ is weakly increasing. In Step 2 of the (only) Lemma in 
the Online Appendix of \citet{Anderlini-Kim-EL:24} we prove that if $b(\cdot)$ is weakly increasing, 
it must be strictly increasing as well. This is sufficient to conclude the proof. \endproof

\section{The Proof of Proposition 
\ref{Proposition: Overbidding}}\label{section app: proof of prn Overbidding}
See the (only) Lemma in the online Appendix to \citet{Anderlini-Kim-EL:24}.

\section{The Proof of Proposition \ref{Proposition: No Limit Inefficiency}}\label{section app: proof of prn No Limit Inefficiency}
Since $\tilde{S}_l \leq S_l^{SA}$ must hold, it suffices to show that $\liminf_{l \to \infty} \tilde{S}_l \geq k$. 
Notice that the total expected payoff among all bidders (excluding the seller) is $\tilde{S}_l - \tilde{R}_l$. Then, observe that
\begin{eqnarray}
    \liminf_{l \to \infty} (\tilde{S}_l-\tilde{R}_l) = \liminf_{l \to \infty} \tilde{S}_l - k.
\end{eqnarray}
Since the expected payoff of each bidder must be nonnegative, we must have $\liminf_{l \to \infty} \tilde{S}_l \geq k$, 
which completes the proof.\endproof

\section{The Proof of Proposition \ref{Proposition: Optimal Auction Revenue}}\label{section app: proof of prn Optimal Revenue}
We begin by establishing that $\limsup_{l \to \infty} R_l^{OA} \leq k$. Recall Myerson's (1981) formula for the expected revenue based on virtual utility.\footnote{This is equation (4.12) in \cite{Myerson:81}} Also recall that virtual utility is always bounded above by the true value. Hence for any fixed $\epsilon>0$ we may write
\begin{eqnarray}
    R_l^{OA} \leq {\rm Prob}[|w_l-k|<\epsilon](k+\epsilon) + (1-{\rm Prob}[|w_l-k|<\epsilon])\bar{w}.
\end{eqnarray}
Then, since $w \xrightarrow{d} k$ and $k$ is a constant, we also have that $\lim_{l \to \infty} {\rm Prob}[|w_l-k|<\epsilon] = 1$. Thus, we may then write
\begin{eqnarray}
    \limsup_{l \to \infty} R_l^{OA} \leq k + \epsilon.
\end{eqnarray}
Since $\epsilon>0$ was arbitrary, we have established $\limsup_{l \to \infty} R_l^{OA} \leq k$.

Then, to conclude the proof 
Proposition \ref{Proposition: Optimal Auction Revenue} 
it suffices to establish the existence of a sequence of auction mechanisms that induces expected revenue 
sequence $\{R_l\}_{l=1}^\infty$ such that $\lim_{l \to \infty} R_l = k$. This is established by Proposition \ref{prn: Approximating Sequence}.\endproof

\section{The Proof of Proposition \ref{Proposition: Tournament Auction Revenue}}\label{section app: proof of prn Tournament Revenue}
Fix any $c \in (0,k)$, and consider the conditional expectation
\begin{eqnarray}
    E[w_l | w_l \leq c] = \int_0^c \Big( 1 - \frac{G_l(\xi)}{G_l(c)} \Big) \,d\xi.
\end{eqnarray}
From the monotonicity of $1-G_l(\xi)/G_l(c)$ with respect to $\xi$, it is clear that $\lim_{l \to \infty} E[w_l | w_l \leq c] = 0$ is equivalent to $\lim_{l \to \infty} G_l(\xi)/G_l(c) = 1$ for all $\xi \in (0,c)$. It then follows that (\ref{eqn: conditional mean converges to zero}) is equivalent to the condition $\lim_{l\to \infty} E[w_l | w_l \leq c] = 0 \; \forall c \in (0,k)$. We will work with the latter condition throughout this proof.

The remainder of the proof proceeds in four steps.

\setcounter{stp}{0}

\begin{step}\label{step: limsup in TA revenue}
$\limsup_{l \to \infty} b_l(v) \leq k$ for all $v \in (0,\bar{v}]$.
\end{step}

By way of contradiction suppose 
that $\limsup_{l \to \infty} b_l(v) > k$ for some $v \in (0,\bar{v}]$. Then, choose a subsequence 
$\{b_{l_m}(v)\}_{m=1}^\infty$ that converges to some $\bar{b}>k$. Along this subsequence, the equilibrium payoff of a first-stage bidder with value $v$ converges to
\begin{eqnarray}
    F(v)^{N-1}[v-k]<0
\end{eqnarray}
which is a contradiction of optimality.

\begin{step}\label{step: subsubsequences in TA revenue}
Suppose that there is some $v_1 \in (0,\bar{v}]$ such that $\lim_{l \to \infty} b_l(v_1) = k$ does not hold (either because the limit does not exist, or because the limit does not equal $k$). Then, we must have $\lim_{l \to \infty} b_l(v) = k$ for all $v \in (v_1,\bar{v}]$.
\end{step}

To prove this claim, first notice that for any $v \in (0,\bar{v}]$, $\lim_{l \to \infty} b_l(v) = k$ does not hold if and only if
\begin{eqnarray}
    \liminf_{l \to \infty} b_l(v) < k.
\end{eqnarray}
This is a direct consequence of Step \ref{step: limsup in TA revenue}.

Hence we begin with the hypothesis that $\liminf_{l \to \infty} b_l(v_1) < k$ for some $v_1 \in (0,\bar{v}]$. Fix any $v_2 \in (v_1,\bar{v}]$. Our goal is to establish $\lim_{l \to \infty} b_l(v_2) = k$.

By our hypothesis there is some subsequence $\{b_{l_m}(v_1)\}_{m=1}^\infty$ such that $\lim_{m \to \infty} b_{l_m}(v_1) = b_1 < k$. We will first show that for any such subsequence, we must have $\liminf_{m \to \infty} b_{l_m}(v_2) \geq k$. Suppose that this is not the case. Then, we can find a finer subsequence $\{b_{l_{m_n}}(v_2)\}_{n=1}^\infty$ such that $\lim_{n \to \infty} b_{l_{m_n}}(v_2) = b_2 < k.$ Along this sub-subsequence, consider the corresponding sub-subsequence of equilibrium payoffs for a first-stage bidder with value $v_1$:
\begin{eqnarray}
    U_{l_{m_n}} \equiv F(v_1)^{N-1} G_{l_{m_n}}(b_{l_{m_n}}(v_1)) \big[v_1 - E[w_{l_{m_n}}|w_{l_{m_n}} \leq b_{l_{m_n}}(v_1)]\big].
\end{eqnarray}
Then, consider also the payoff of a first-stage bidder with true value $v_1$ but bidding according to $b_{l_{m_n}}(v_2)$. We would then have
\begin{eqnarray}
\begin{split}
    \hat{U}_{l_{m_n}} & = F(v_2)^{N-1} G_{l_{m_n}}(b_{l_{m_n}}(v_2))\big[v_1 - E[w_{l_{m_n}}|w_{l_{m_n}} \leq b_{l_{m_n}}(v_2)]\big] \\ & \geq F(v_2)^{N-1} G_{l_{m_n}}(b_{l_{m_n}}(v_1))\big[v_1 - E[w_{l_{m_n}}|w_{l_{m_n}} \leq b_{l_{m_n}}(v_2)]\big].
\end{split}
\end{eqnarray}
Then, observe that
\begin{eqnarray}
\begin{split}
    \frac{\hat{U}_{l_{m_n}}}{U_{l_{m_n}}} & \geq \bigg(\frac{F(v_2)}{F(v_1)}\bigg)^{N-1} \frac{v_1-E[w_{l_{m_n}}|w_{l_{m_n}}\leq b_{l_{m_n}}(v_2)]}{v_1-E[w_{l_{m_n}}|w_{l_{m_n}}\leq b_{l_{m_n}}(v_1)]} \\
    & \rightarrow \bigg(\frac{F(v_2)}{F(v_1)}\bigg)^{N-1} > 1
\end{split}
\end{eqnarray}
as $n\to \infty$. However, this clearly contradicts the optimality of the equilibrium bid given $v_1$. Therefore, we have established our claim that $\lim_{m \to \infty} b_{l_m}(v_1)<k$ implies $\liminf_{m \to \infty} b_{l_m}(v_2) \geq k$.

We now establish the main claim of Step \ref{step: subsubsequences in TA revenue}. Suppose for contradiction that $\liminf_{l \to \infty} b_l(v_2) < k$ 
so that there exists some subsequence $\{b_{l_m}(v_2)\}_{m=1}^\infty$ such that $\lim_{m \to \infty} b_{l_m}(v_2) < k$. 
Then, it is clear from the monotonicity of the bidding function that there is a finer subsequence $\{b_{l_{m_n}}(v_1)\}_{n=1}^\infty$ 
given $v_1$ such that the sub-subsequence converges to a bid strictly below $k$. However, we then know from the claim established above 
that $\liminf_{n \to \infty} b_{l_{m_n}}(v_2) \geq k,$ which is clearly a contradiction.

\begin{step}\label{step: lim = k in proof of TA}
$\lim_{l \to \infty} b_l(v) = k$ for all $v \in (0, \bar{v}]$.
\end{step}

From Step \ref{step: subsubsequences in TA revenue}, it is clear that there is at most one $v^* \in (0, \bar{v}]$ for which $\lim_{l \to \infty} b_l(v^*) = k$ does not hold. However, we can always choose any $\tilde{v} \in (0,v^*)$, for which we are sure that the convergence condition $\lim_{l \to \infty} b_l(\tilde{v}) = k$ holds. Then, from the monotonicity of the bidding function and the result of Step 1, it is clear that $\lim_{l \to \infty} b_l(v^*) = k$ must actually hold. Thus, Step 3 is established.

\begin{step}
 We now establish Proposition \ref{Proposition: Tournament Auction Revenue}.
\end{step}

For any $v \in (0, \bar{v}]$, we clearly have
\begin{eqnarray}
    R_l^{TA} \geq \{1-F(v)^N\} E[\min\{b_l(v),w_l\}].
\end{eqnarray}

Then, observe that for any fixed $\epsilon>0$,
\begin{eqnarray}
    \lim_{l \to \infty} {\rm Prob}[\min\{b_l(v),w_l\} \geq k - \epsilon] = \lim_{l \to \infty} [1-G_l(k-\epsilon)] = 1 
\end{eqnarray}
due to Step \ref{step: lim = k in proof of TA}.

Hence for any $\epsilon>0$ and $v \in (0,\bar{v}]$ we have 
\begin{eqnarray}
    R_l^{TA} \geq \{1-F(v)^N\}(k-\epsilon) \cdot {\rm Prob}[\min\{b_l(v),w_l\} \geq k - \epsilon]
\end{eqnarray}
\begin{eqnarray}
    \liminf_{l\to \infty} R_l^{TA} \geq \{1-F(v)^N\}(k-\epsilon).
\end{eqnarray}
Hence it is clear that
\begin{eqnarray}
    \liminf_{l\to \infty} R_l^{TA} \geq k.
\end{eqnarray}
Then, from Proposition \ref{Proposition: Optimal Auction Revenue}, it is clear that
\begin{eqnarray}    
    \lim_{l \to \infty} R_l^{TA} = k.
\end{eqnarray}
\endproof

\section{The Proof of Proposition \ref{prn: Approximating Sequence}}\label{section app: proof of Approximating Sequence}
We prove the first claim of existence by direct construction. Fix any $\epsilon>0$ such that $k-\epsilon>0$. We can then construct a sequence $\{L_n\}_{n=1}^\infty \subseteq \mathbb{N}$ that is strictly increasing based on the following procedure. First define $L_0 \equiv 0$. Then, for each $n \in \mathbb{N}$, we know that since $\lim_{l \to \infty} G_l(k-\epsilon/n) = 0$, there is some $L_n \in \mathbb{N}$ with $L_n > L_{n-1}$ such that $G_l(k-\epsilon/n) \leq 1/n$ for all $l \geq L_n$.

Now we define $\{r_l\}_{l=1}^\infty$ as below:
\begin{eqnarray}
    r_l = k - \frac{\epsilon}{n} \quad {\rm for} \; l = L_n,L_n + 1, ..., L_{n+1}-1, \; n = 1,2,...
\end{eqnarray}
Then, by construction we must have that for each $n \in \mathbb{N}$
\begin{eqnarray}
    R_l^{SA}(r_l) \geq (1-G_l(r_l))r_l \geq \Big(1-\frac{1}{n}\Big)\Big(k - \frac{\epsilon}{n}\Big) \quad \forall \; l \geq L_n.
\end{eqnarray}
Therefore, we must then have that
\begin{eqnarray}
    \liminf_{l \to \infty} R_l^{SA}(r_l) \geq k.
\end{eqnarray}
In light of the inequality $\limsup_{l \to \infty} R_l^{OA} \leq k$ established in the proof of Proposition \ref{Proposition: Optimal Auction Revenue}, the first claim of existence is then proven.

We now establish the second claim of Proposition \ref{prn: Approximating Sequence}. First observe that given a reserve price $r_l<k$, the revenue is certainly bounded above by $\max \{r_l,\bar{v}\}$. From this it is clear that if $R_l^{SA}(r_l) \rightarrow k$, then $\liminf_{l \to \infty} r_l \geq k$. Furthermore, in light of Proposition \ref{prn: Overshooting}, it is clear that if $R_l^{SA}(r_l) \rightarrow k$, then $\limsup_{l \to \infty} r_l \leq k$. This then establishes our second claim.\endproof

\section{The Proof of Proposition \ref{prn: Undershooting}}\label{section app: proof of Undershooting}
We begin by noting that for $r_l \geq \bar{v}$
\begin{eqnarray}\label{equation: revenue of second price with reserve}
    R_l^{SA}(r_l) = G_l(r_l) E[v_{2:N}] + (1-G_l(r_l))r_l
\end{eqnarray}
and
\begin{eqnarray}\label{equation: surplus of second price with reserve}
    S_l^{SA}(r_l) = G_l(r_l) E[v_{1:N}] + (1-G_l(r_l))E[w_l|w_l \geq r_l]
\end{eqnarray}

Clearly $\lim_{l \to \infty} G_l(r_l) = 0$ and $\lim_{l \to \infty} E[w_l|w_l \geq r_l] = k$. Then, taking the limit 
as $l \to \infty$ of (\ref{equation: revenue of second price with reserve}) and 
(\ref{equation: surplus of second price with reserve}), we obtain the desired results.\endproof

\section{The Proof of Proposition \ref{prn: Overshooting}}\label{section app: proof of Overshooting}

Notice that $\lim_{l \to \infty} G_l(r_l) = 1$. Then, taking the limit as $l \to \infty$ of (\ref{equation: revenue of second price with reserve}) 
and (\ref{equation: surplus of second price with reserve}), we obtain the desired results.\endproof

\section{The Proof of Proposition \ref{prn: Convergence from Below}}\label{section app: proof of Convergence from Below}
Convergence in distribution as in (\ref{eqn: convergence in distribution}) implies that $G_l(w) \rightarrow 0$ if $w<k$ and $G_l(w) \rightarrow 1$ if $w>k$. However, when $w=k$, the asymptotic behavior of $G_l(k)$ is indeterminate. We can have $G_l(w) \rightarrow p$ for any $p \in [0,1]$. Let us begin by fixing any such $p \in [0,1]$.

Let us first define $r_0 \equiv 0$. Then, for each $l=1,2,...,$ we have from the continuity of $G_l$ that there exists $r_l \in (r_{l-1},k)$ such that 
\begin{eqnarray}
   G_l(k)-\frac{1}{l} < G_l(r_l) < G_l(k).
\end{eqnarray}
It is immediately clear from this construction that we must then that $r_l \uparrow k$ with $r_l < k$ for each $l$. Furthermore, we must have
\begin{eqnarray}\label{equation: convergence to p}
    \lim_{l \rightarrow \infty} G_l(r_l) = \lim_{l \to \infty} G_l(k) = p.
\end{eqnarray}

Let us then recall (\ref{equation: revenue of second price with reserve}) 
and (\ref{equation: surplus of second price with reserve}) as formulas for the expected revenue and surplus for the second price auction with reserve price $r_l$. Firstly, in light of (\ref{equation: convergence to p}) and $r_l \uparrow k$, it follows from equation (\ref{equation: revenue of second price with reserve}) that we must have
\begin{eqnarray}
    \lim_{l \to \infty} R_l^{SA}(r_l) = p E[v_{2:N}] + (1-p) k.
\end{eqnarray}

Let us then consider the limit of the expected surplus as in (\ref{equation: surplus of second price with reserve}). We will consider two separate cases:

\underline{Case 1: $p = 1$.}

Notice that $E[w_l | w_l \geq r_l]$ is bounded in $[0,\bar{w}]$ across all $l$. Thus, it follows that
\begin{eqnarray}
    \lim_{l \to \infty} (1-G_l(r_l)) E[w_l | w_l \geq r_l] = 0
\end{eqnarray}
under the case of $p = 1$. We then have from taking the limit of (\ref{equation: surplus of second price with reserve}) that
\begin{eqnarray}
    \lim_{l \to \infty} S_l^{SA}(r_l) = E[v_{1:N}],
\end{eqnarray}
which is our desired conclusion.

\underline{Case 2: $p \in [0,1)$.}

Let us consider the conditional distribution of $w_l$ given $w_l \geq r_l$. The support is $[r_l,\bar{w}]$, and the conditional distribution is given by
\begin{eqnarray}
    {\rm Prob}[w_l \leq \xi | w_l \geq r_l ] = \frac{G_l(\xi) - G_l(r_l)}{1- G_l(r_l)}
\end{eqnarray}
for all $\xi \in [r_l,\bar{w}]$. Then, we may write the conditional expectation as
\begin{eqnarray}
    E[w_l | w_l \geq r_l] = r_l + \int_{r_l}^{\bar{w}} 1 - \frac{G_l(\xi) - G_l(r_l)}{1- G_l(r_l)} \, d\xi.
\end{eqnarray}
Then, using the dominated convergence theorem and $r_l \uparrow k$, we have that
\begin{eqnarray}
    \lim_{l \to \infty} E[w_l | w_l \geq r_l] = k + \int_{k}^{\bar{w}} 1 - \frac{1 - p}{1- p} \, d\xi = k.
\end{eqnarray}
Therefore, taking the limit of (\ref{equation: surplus of second price with reserve}), we obtain
\begin{eqnarray}
    \lim_{l \to \infty} S_l^{SA}(r_l) = p E[v_{1:N}] + (1-p)k,
\end{eqnarray}
which completes the proof.\endproof

\end{appendix}
\end{document}